\documentclass[twocolumn,showpacs,superscriptaddress,a4paper,amsmath,amssymb]{revtex4}

\usepackage{graphicx}
\usepackage{dcolumn}
\usepackage{bm}
\usepackage{siunitx}
\usepackage{graphicx}
\usepackage{graphics}
\usepackage{amssymb}
\usepackage{amsmath}
\usepackage{epsfig}
\usepackage{latexsym}
\usepackage{color}
\usepackage[utf8]{inputenc}
\usepackage{amsfonts}
\usepackage{amsmath}
\usepackage{amssymb}
\usepackage{color}
\usepackage{float}
\usepackage{textcomp}
\usepackage{subfigure}
\usepackage{graphicx}
\usepackage{tikz}
\usepackage{circuitikz}
\usepackage{pdfpages}
\usepackage{graphicx}
\usepackage[export]{adjustbox}
\usepackage{rotating}
\usepackage{subfigure}
\usepackage{cleveref}
\usepackage{fontenc}
\usepackage{cancel}

\newcommand{\hide}[1]{\relax}

\newcommand{\Og}{\ensuremath{\Omega}}

\newcommand{\OLC}{\ensuremath{\Omega_\mathrm{LC}}}

\newcommand{\GLC}{\ensuremath{\Gamma_\mathrm{LC}}}

\newcommand{\chiLC}{\chi_\text{LC}}

\newcommand{\bx}{\ensuremath{\bar x}}
\newcommand{\bq}{\ensuremath{\bar q}}

\newcommand{\mhat}{}

\newcommand{\dhx}{\ensuremath{\delta \mhat x}}

\newcommand{\dhp}{\ensuremath{\delta \mhat p}}
\newcommand{\dhq}{\ensuremath{\delta \mhat q}}
\newcommand{\dhphi}{\ensuremath{\delta \mhat \phi}}

\newcommand{\dhV}{\ensuremath{\delta\! \mhat V}}

\newcommand{\new}[1]{\textcolor{black}{#1}}

\begin{document}
\setlength{\abovedisplayskip}{3pt}
\setlength{\belowdisplayskip}{3pt}
\title{
	Sensitivity--bandwidth limit in a multi--mode opto--electro--mechanical transducer
	}

\author{I. Moaddel Haghighi}
\affiliation{\text{School of Science and Technology, Physics Division, University of Camerino, I-62032 Camerino (MC), Italy}}
\author{N. Malossi}
\affiliation{\text{School of Science and Technology, Physics Division, University of Camerino, I-62032 Camerino (MC), Italy}}
\affiliation{INFN, Sezione di Perugia, Italy}
\author{R. Natali}
\affiliation{\text{School of Science and Technology, Physics Division, University of Camerino, I-62032 Camerino (MC), Italy}}
\affiliation{INFN, Sezione di Perugia, Italy}
\author{G. Di Giuseppe}
\affiliation{\text{School of Science and Technology, Physics Division, University of Camerino, I-62032 Camerino (MC), Italy}}
\affiliation{INFN, Sezione di Perugia, Italy}
\author{D. Vitali}
\affiliation{\text{School of Science and Technology, Physics Division, University of Camerino, I-62032 Camerino (MC), Italy}}
\affiliation{INFN, Sezione di Perugia, Italy}
\affiliation{CNR-INO, L.go Enrico Fermi 6, I-50125 Firenze, Italy}

\date{\today}

\begin{abstract}
An opto--electro--mechanical system formed by a nanomembrane capacitively coupled to an LC resonator and to an optical interferometer has been recently employed for the high--sensitive optical readout of radio frequency (RF) signals [T. Bagci, \emph{et~al.}, Nature {\bf 507}, 81 (2013)]. Here we propose and experimentally demonstrate how the bandwidth of such kind of transducer can be increased by controlling the interference between two--electromechanical interaction pathways of a two--mode mechanical system. With a proof--of--principle device
\new{operating at room temperature, we achieve a sensitivity of $\SI{300}{\nano\volt\per\sqrt{\hertz}}$ over a bandwidth of 15 kHz in the presence of radiofrequency noise, and an optimal shot-noise limited sensitivity of $\SI{10}{\nano\volt\per\sqrt{\hertz}}$ over a bandwidth of 5 kHz. We discuss strategies for improving the performance of the device, showing that, for the same given sensitivity, a mechanical multi--mode transducer can achieve a bandwidth}
significantly larger than that of a single-mode one.
\end{abstract}

\maketitle

\section{Introduction}

Optomechanical and electromechanical systems have recently shown an impressive development~\cite{RMP2014},
and lately entered a quantum regime in which quantum states of nanogram--size mechanical resonators and/or electromagnetic fields have been generated and manipulated~\cite{OConnell2010,Teufel2011a,Chan2011,Verhagen2012,Safavi-Naeini2013,Palomaki2013a,Peterson2016,Clark2016,Riedinger2016}. They have been also suggested and already employed for testing fundamental theories~\cite{Pikovski2012,Bawaj2015}, and for quantum--limited sensing~\cite{Chaste2012,Moller2017}. Furthermore, nanomechanical resonators can be simultaneously coupled to a large variety of different degrees of freedom and therefore they can transduce signals at disparate frequencies with high efficiency~\cite{Stannigel2010, Regal2011,Taylor2011,Wang2012,Tian2012,Barzanjeh2012}, either in the classical and in the quantum domain. Reversible transduction between optical and radio--frequency/microwave (RF/MW) signals is nowadays particularly relevant, both in classical and quantum communication systems, and first promising demonstrations with classical signals have been recently achieved~\cite{Bagci2013,Andrews2014,Fink2016}.
In particular the conversion of RF/MW signals into optical ones can be exploited for the high--sensitive detection of weak RF/MW signals, by taking advantage of the fact that the homodyne detection of laser light can be quantum noise limited with near--unit quantum efficiency. This could avoid many of the noise sources present for low frequency signals and could be useful for example in radio astronomy, medical imaging, navigation, and classical and quantum communication. Bagci~\emph{et~al.}~\cite{Bagci2013} reported a first important demonstration of this idea with an optical interferometric detection of RF signals with $\SI{800}{\pico\volt\per\sqrt{Hz}}$ sensitivity, and which could be improved down to $\SI{5}{\pico\volt\per\sqrt{Hz}}$ in the limit of strong electromechanical coupling. In this device, weak RF signals drive an LC resonator quasi--resonantly interacting with the nano--mechanical transducer, whose motion induces an optical phase shift which is then detected with quantum--limited sensitivity.
\new{A first application of an opto-electro-mechanical transducer for nuclear magnetic resonance detection has been recently demonstrated \cite{NMR}}.

\begin{figure}[t!]
\includegraphics[width=0.475\textwidth]{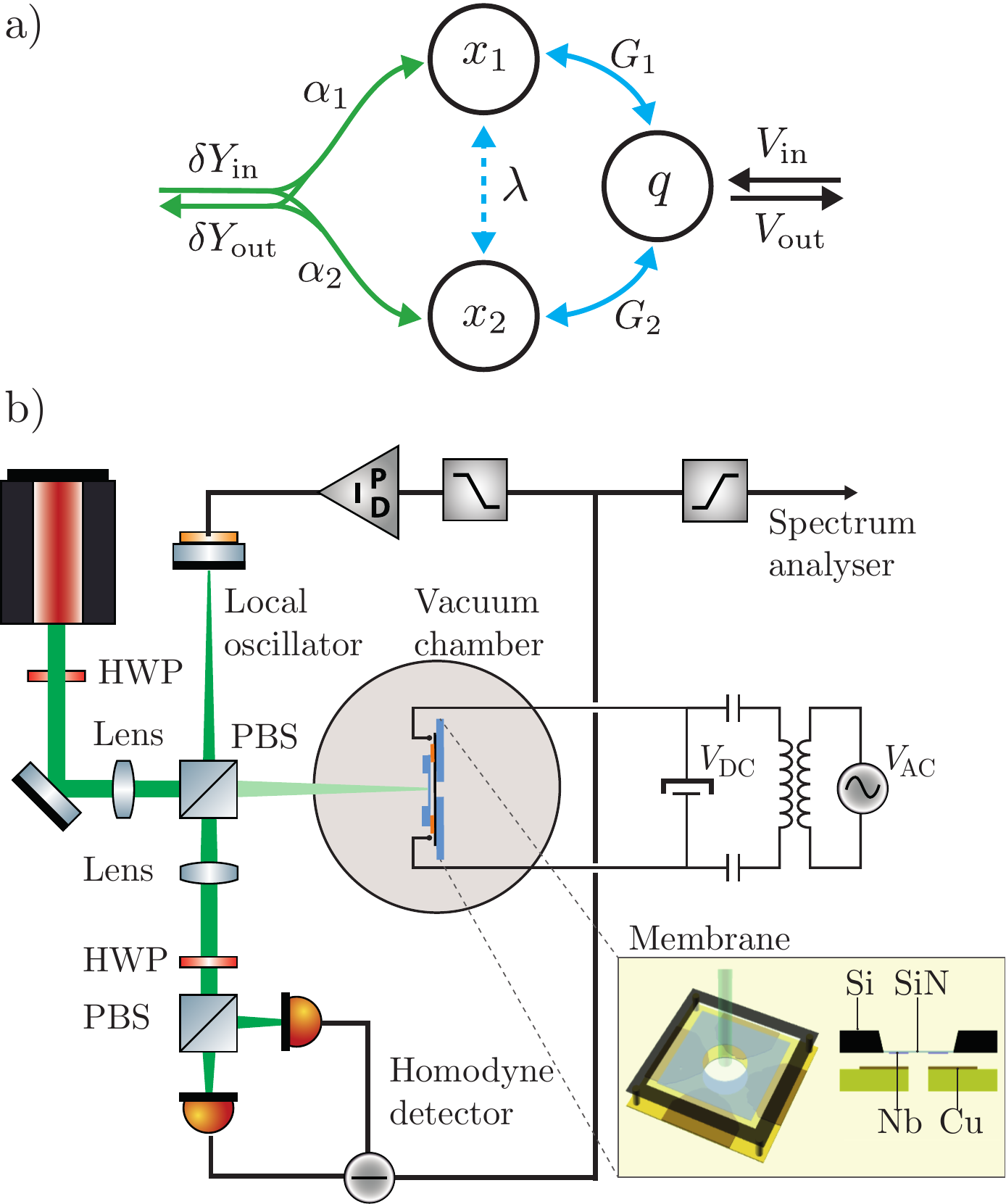}
\caption{
	{\bf a)} Scheme of the RF--to--optical transducer and of the interference between the two transduction pathways through the two mechanical modes, $x_1$, and $ x_2$. The two modes are simultaneously capacitively coupled with electromechanical coupling $G_1$ and $G_2$ to the same LC resonator $q$, and eventually through a direct mechanical interaction, $\lambda$. At the same time the motion of the two resonators is readout by an optical interferometer using the light reflected from the membrane, $\delta Y_\mathrm{out}$, with optical couplings $\alpha_1$ and $\alpha_2$. Since modulation of the phase noise of the optical beam occurs through two different paths (via mode 1 or mode 2), the signal detected by the optical interferometer depends upon the interference between these two paths, which in turn can be controlled through the electrode configuration of the membrane capacitor.
	{\bf b)} Experimental setup. An RF resonator is constituted by an inductor and a membrane capacitor placed in a vacuum chamber evacuated at $\SI{1e-7}{\milli\bar}$. The mechanical displacement is revealed by homodyne detection of the light reflected by the membrane. The electromechanical coupling is controlled by applying a dc--bias $V_\mathrm{DC}$ over two electrodes. The system is driven inductively, through two capacitors, with an RF signal $V_\mathrm{AC}$ using an antenna. Inset:  a $1\si{\milli\meter}\times 1\si{\milli\meter}$ SiN membrane coated with a \SI{27}{\nano\meter} Nb film stands on top of four segment electrodes, forming a capacitor $C_\mathrm{m}(x)$ modulated by the membrane motion.}
	\label{fig:1}
 \end{figure}

In Ref.~\cite{Bagci2013} the detection bandwidth depends upon the LC bandwidth and the electromechanical coupling \cite{Taylor2011,Bagci2013}, and is an important figure of merit of such kind of transducers \cite{Zeuthen2016}. Finding systematic ways of increasing the detection bandwidth is of fundamental importance in many of the above--mentioned applications: for example, more radio-astronomical sources could be detected, while in communication networks RF signals could be faster detected and processed.
Here we show with a proof-of-principle experiment that a viable way to increase the bandwidth of opto--electromechanical transducers is to couple the LC circuit simultaneously to two (or more) mechanical modes with nearby frequencies, and suitably engineer the two electromechanical couplings in order to realise a constructive interference between the two RF--to--optical signal transductions mediated by each mechanical mode [see Fig.~\ref{fig:1}(a)].
\new{Multimode optomechanical \cite{Nielsen2017} and electromechanical \cite{Massel2012,Noguchi2016,Han2016} systems have been recently studied and operated in a quantum regime, but here we exploit them with the aim of improving bandwidth and sensitivity of an opto-electro-mechanical transducer}.
The mechanical transducer is a $\SI{1x1}{\milli\meter}$ SiN membrane of $\SI{50}{\nano\meter}$ thickness, coated with a $\SI{27}{\nano\meter}$ Nb film with a central circular hole (Norcada Inc., see inset of Fig.~\ref{fig:1}), capacitively coupled through Cu electrodes to an LC resonator and operated at room temperature. The mechanical modes exploited are the split doublet $(1,2)-(2,1)$ revealed by optical homodyne detection.
\new{The achieved sensitivity of the two--mode transducer is $\SI{300}{\nano\volt\per\sqrt{\hertz}}$ over a bandwidth of 15 kHz in the presence of RF noise, and the optimal shot--noise limited sensitivity is}
$\SI{10}{\nano\volt\per\sqrt{\hertz}}$ over a bandwidth of 5 kHz. The sensitivities are obtained in the case of electromechanical couplings for the two modes equal to $G_1=\SI{118.41}{\volt\per\meter}$ and $G_2=\SI{-115.31}{\volt\per\meter}$.
However, as we show in Sec. \ref{theory}, the method is general and could be exploited to reach larger bandwidths at a sensitivity comparable to that of single mode transducers~\cite{Bagci2013}.
The paper is organized as follows. In Sec. \ref{theory} we introduce the multi-mode transducer theoretical framework. In Sec. \ref{sec:Experiment} we show and discuss the experimental result showing the performance of our device, and we also see how one can improve the design so that a two-mode transducer can achieve a larger bandwidth at the same sensitivity of a single-mode electromechanical transducer. Concluding remarks are provided in Sec. \ref{Concl}.

\section{Theoretical Framework}\label{theory}
The system studied here is formed by a nanomechanical system capacitively coupled to an LC resonator.
\new{We generalize here to the multi-mode case the treatment of Ref. \cite{Bagci2013}}.
The nanomechanical system has a number of vibrational normal modes which can be described in terms of effective mechanical resonators with mass $m_i$, frequency $\omega_i$, displacement $x_i$, momentum $p_i$, so that the effective Hamiltonian of the system is
\begin{equation}
   H=\sum_i\frac{p_i^2}{2 m_i}+\frac{m_i \omega_{i}^{2} x_i^2}{2}+\frac{\phi^2}{2 L}+\frac{q^2}{2 C(\left\{x_i\right\})}-q V,
\end{equation}
where $\phi$ is the flux in the inductor, $q$ is the charge on the capacitors, and $V$ is the voltage bias across the capacitor.
The coupling arises due to the displacement dependence of the capacitance $C(\left\{x_i\right\})$. This Hamiltonian directly leads to the Langevin equations
\begin{align}
  \dot x_i &=\frac{p_i}{m_i},\\
  \dot p_i &= -m_i \omega_i^2 x_i - \frac{q^2}{2} \frac{\partial}{\partial x_i} \left(\frac{1}{C(\left\{x_i\right\})}\right)- \Gamma_i p_i + F_i,\\
  \dot q &= \frac{\phi}{L},\\
  \dot \phi &= -\frac{q}{C(x)}-\GLC \phi+V,
\end{align}
in which the terms corresponding to the damping rate $\Gamma_i$ of the $i$--th membrane mode, and to the resistive dissipation rate $\Gamma_\mathrm{LC}=R/L$ of the LC resonant circuit have been included, as well as driving forces $F_i$ acting on each membrane mode.  Assuming that $F_i$ are zero--mean thermal Langevin forces, and writing the applied voltage as a large d.c.\ offset and a small fluctuating input
\begin{eqnarray}
   V(t) &=& V_\mathrm{DC}+ \delta V(t),
\end{eqnarray}
we can linearize the Langevin equations around an equilibrium state of the system characterized by $(\bar x_i, \bar p_i, \bar q, \bar \phi)$ and satisfying the conditions
\begin{align}
 m_{i} \omega_i^2 \bx_i
  	&= - \frac{\bq^2}{2}  \left. \frac{\partial}{\partial x_i} \left(\frac{1}{C(\left\{x_i\right\})}\right)\right|_{x_i=\bx_i}\\ \nonumber
  	&=  \frac{\bq^2}{2}\left. \frac{\partial C(\left\{x_i\right\})}{\partial x_i}\right|_{x_i=\bx_i}\frac{1}{C(\left\{\bar{x}_i\right\})^2},\\
  \bq &= V_\mathrm{DC} C(\left\{\bar{x}_i\right\}),\\
  \bar p_i &= \bar \phi = 0.
\end{align}
The dynamical equations for the small fluctuations, provided that the system is stable, are given by

\begin{align}
  \delta\dot x_i(t) &= \frac{\dhp_i(t)}{m_i},\\
  \delta \dot p_i(t) &= -m_i \omega_i^2 \dhx_i(t)  \\
  	&- \underbrace{\frac{\bq^2}{2} \left.  \frac{\partial^2}{\partial^2 x_i} \left(\frac{1}{C(\left\{x_i\right\})}\right)\right|_{x_i=\bx_i}}_{2 m \omega_i
 \,\Delta\omega_i} \dhx_i(t)  \nonumber \\
 &- \underbrace{\frac{\bq^2}{2} \left. \sum_{j \neq i} \frac{\partial^2}{\partial x_i \partial x_j} \left(\frac{1}{C(\left\{x_i\right\})}\right)\right|_{x_i=\bx_i}}_{\lambda_{ij}} \dhx_j(t) \nonumber  \nonumber \\
  	&- \Gamma_i \delta p_i - \underbrace{\bq \left. \frac{\partial}{\partial x_i} \left(\frac{1}{C(\left\{x_i\right\})}\right)\right|_{x_i=\bx_i}}_{G_i}  \dhq(t)+ F_i,  \nonumber \\
  \delta \dot q(t) &= \frac{\dhphi(t)}{L}, \\
  \delta\dot{\phi}(t)  &=- \frac{\dhq(t)}{C(\left\{\bar{x}_i\right\})}  - \GLC \dhphi(t)+ \delta V(t)\nonumber \\
	&- \sum_j \underbrace{\bq \left. \frac{\partial}{\partial x_j} \left(\frac{1}{C(\left\{x_i\right\})}\right)\right|_{x_i=\bx_i}}_{G_j} \dhx_j(t)  .
\end{align}
Here, we have introduced the electro--mechanical coupling parameters
\begin{eqnarray}\label{e:G_def}
  G_i	&=& \bq \left. \frac{\partial}{\partial x_j} \left(\frac{1}{C(\left\{x_i\right\})}\right)\right|_{x_i=\bx_i},
\end{eqnarray}
the mechanical coupling between the vibrational normal modes induced by the second-order dependence of the capacitance upon the membrane deformation,
\begin{eqnarray}\label{e:lam_def}
  \lambda_{ij}	&=& \frac{\bq^2}{2} \left. \sum_{j \neq i} \frac{\partial^2}{\partial x_i \partial x_j} \left(\frac{1}{C(\left\{x_i\right\})}\right)\right|_{x_i=\bx_i},
\end{eqnarray}
and the mechanical frequency shifts
\begin{eqnarray}\label{freqshi}
  \Delta\omega_i &=& \frac{\bq^2}{4 m_i \omega_i} \left. \frac{\partial^2}{\partial x_i^2} \left(\frac{1}{C(\left\{x_i\right\})}\right)\right|_{x_i=\bx_i}.
\end{eqnarray}
Absorbing the frequency shifts into re-defined $\omega_i$ and transforming to the Fourier domain yields
\begin{eqnarray}
- i \Og \, \dhx_i(\Og) & =& \dhp_i(\Og) /m_i, \label{e:eomlfx} \\
- i \Og \, \dhp_i(\Og) & =& - m_i \omega_i^2 \,\dhx_i(\Og) -\sum_j \lambda_{ij}\,\dhx_j(\Og) \nonumber \\
&-& \Gamma_i \,\dhp_i(\Og) - G_i \,\dhq(\Og) +F_i(\Og) ,	\label{e:eomlfp}\\
- i \Og \, \dhq(\Og)  & =& \dhphi(\Og) / L 	,\label{e:eomlfq}\\
- i \Og \, \dhphi(\Og) & =& -\dhq(\Og)/C- \GLC \, \dhphi(\Og)+ \dhV(\Og)  \nonumber \\
 &-&  \sum_j G_j \, \dhx_j(\Og). \label{e:eomlfphi}
\end{eqnarray}
These algebraic equations can be used to calculate the response of the system to excitations through a force or voltage drive.
For notational convenience, we define the susceptibilities
\begin{eqnarray}
  \chi_i(\Og)&=& \frac{1}{m_i\left(\omega_i^2-\Og^2-i \Og \Gamma_i\right) },\\
  \chiLC(\Og)&=& \frac{1}{L\left(\OLC^2-\Og^2-i \Og \GLC\right)}. \label{suscLC}
\end{eqnarray}
of the $i$--th mechanical mode and of the LC resonator, respectively, and we have defined the circuit resonance frequency $\OLC = \left(LC\right)^{-1/2}$.
In the general case of many membrane modes the solution can be easily derived when $\lambda_{ij}=0$, i.e., without the direct mechanical coupling mediated by the capacitance and in the presence of only the indirect coupling through the LC resonator.
\subsection{Two mechanical modes}
We restrict now to the case of our system where the detection bandwidth includes only two mechanical modes and the effect of the other spectator modes is negligible (i.e., it falls below the noise level). Using Eqs. (\ref{e:eomlfx})--(\ref{suscLC}) one can write
\begin{align}
\chi_1(\Og)^{-1} \dhx_1(\Og) &=-\lambda  \dhx_2(\Og) - G_1 \, \dhq(\Og)
+ F_1(\Og),	 \label{e:1m}\\
\chi_2(\Og)^{-1} \dhx_2(\Og) &=-\lambda  \dhx_1(\Og) - G_2 \, \dhq(\Og) + F_2(\Og), \label{e:2m}\\
\chiLC(\Og)^{-1} \dhq(\Og) &= - G_1 \dhx_1(\Og)-G_2 \dhx_2(\Og)+ \dhV(\Og). 	\label{e:lc}
\end{align}
From the last equation we have
\begin{align}
   \delta q(\Omega) = -\chi_{LC}(\Omega)\, [G_1 \delta x_1(\Omega)+G_2 \delta x_2(\Omega) - \delta V(\Omega)]\,,
\end{align}
and substituting in the first two we get
\begin{align}
	\xi_1\,\delta x_1(\Omega) &= \beta\,\delta x_2(\Omega) - G_1\,\chi_{LC}\, \delta V(\Omega)+F_1(\Omega),\\
	\xi_2\,\delta x_2(\Omega) &= \beta\,\delta x_1(\Omega) - G_2\,\chi_{LC}\, \delta V(\Omega)+F_2(\Omega),
\end{align}
with $\xi_i = \chi_i(\Omega)^{-1} - G_i^2\,\chi_{LC}(\Omega)$ and $\beta = [G_1G_2\,\chi_{LC}(\Omega)-\lambda]$\,.
Then we have ($i=1,2$)
\begin{eqnarray}\label{deltaxinter}
	\delta x_i(\Omega) &=& \frac{\xi_{3-i}\,F_i(\Omega)}{\xi_1\xi_2 - \beta^2} +
					\frac{\beta\,F_{3-i}(\Omega) }{\xi_1\xi_2 - \beta^2} \nonumber \\
	&-& \frac{\xi_{3-1}G_i + \beta G_{3-i}}{\xi_1\xi_2 - \beta^2}\chi_{LC}(\Omega) \delta V(\Omega)\,,
\end{eqnarray}
The signal detected by the optical interferometer is the phase quadrature $\delta Y_{\rm out}$ of the light reflected from the membrane, which can be written in the frequency domain as
\begin{eqnarray} \label{eq:detected}
\delta Y_{\rm out}(\Og)= \alpha_1 \dhx_1(\Og)+\alpha_2 \dhx_2(\Og)+\delta Y_{\rm in}(\Og),
\end{eqnarray}
that is, it is the sum of the vacuum phase noise, $\delta Y_{\rm in}(\Og)$, and the displacement fluctuations of the two mechanical modes weighted by the optomechanical couplings $\alpha_i$, which depend upon the overlap of the selected membrane mode with the transverse profile of the optical field.
\new{We calibrate the output signal as a displacement spectrum, so that $\delta Y(\Og)$ has the same units of $\delta x_j(\omega)$, that is, $\mathrm{m/\sqrt{Hz}}$. As a consequence the couplings $\alpha_j$ coincide with the dimensionless transverse overlap parameters defined in Eq. (A6) (see Appendix A)}.
{Using the fact that the four noises, $F_1$, $F_2$, $\delta V$ and $\delta Y_{\rm in}$ are uncorrelated, we can write the output optical phase spectrum as the sum of four independent terms}
\begin{widetext}
\begin{align}\label{spegia}
	S_{\rm out}(\Omega) =
		   \left|\frac{\alpha_1 \xi_2 + \alpha_2 \beta}{\xi_1\xi_2 - \beta^2}\right|^2 S_{F1}(\Omega)
		&+ \left|\frac{\alpha_2 \xi_1 + \alpha_1 \beta}{\xi_1\xi_2 - \beta^2}\right|^2 S_{F2}(\Omega)
		+ S_{\rm in}(\Omega)
					\nonumber \\
		&+ \left|\frac{\alpha_1(\xi_2G_1 + \beta G_2) + \alpha_2(\xi_1G_2 + \beta G_1)}{\xi_1\xi_2 - \beta^2}\right|^2 \left|				\chi_{LC}(\Omega)\right|^2 S_{\delta V}(\Omega)
		,
\end{align}
\end{widetext}
where $S_{Fj}(\Omega)=2m_j\Gamma_j k_B T$, $j=1,2$ are the Brownian force noise spectra, with $T$ the system temperature, $S_{\delta V}(\Omega)$  is the noise voltage at the input of the LC circuit, and $S_{\rm in}(\Omega)$ is the optical shot noise spectrum. Let us now try to readjust and rewrite this general expression for the detected spectrum in order to get some physical intuition from it. We first define the effective mechanical susceptibilities of the two modes, modified by the interaction with the LC circuit,  ($i=1,2$)
\begin{align}
\chi_{i}^{\rm eff}(\Omega) &= \frac{\xi_{3-i}}{\xi_1\xi_2 - \beta^2}, \\
\left[\chi_{i}^{\rm eff}(\Omega)\right]^{-1} &=  \chi_{i}^{-1}(\Omega) -G_i^2\chi_{\rm LC}(\Omega)\nonumber \\
&- \frac{\beta^2}{\chi_{3-i}^{-1}(\Omega)-G_{3-i}^2\chi_{\rm LC}(\Omega)} .
\end{align}
The detected spectrum of Eq. (\ref{spegia}) can be then rewritten as
\begin{align}\label{spegia2}
S_{\rm out}(\Omega) & = \left|\alpha_1  + \alpha_2 \mu_2(\Omega)\right|^2 \left |
\chi_{1}^{\rm eff}(\Omega)\right|^2 \,S_{F1}(\Omega) \nonumber \\
& +\left|\alpha_2 + \alpha_1 \mu_1(\Omega)\right|^2 \left | \chi_{2}^{\rm eff}(\Omega)\right|^2 \,S_{F2}(\Omega) \\
&+ \left|I(\Omega)\right|^2 \left| \chi_{LC}(\Omega)\right|^2  S_{\delta V}(\Omega)+ S_{\rm in}(\Omega), \nonumber
\end{align}
where
\begin{align}
 I(\Omega)& =\alpha_1 \chi_{1}^{\rm eff}(\Omega)\left[G_1+G_2 \mu_2(\Omega)\right]
 \nonumber \\
& + \alpha_2 \chi_{2}^{\rm eff}(\Omega)\left[G_2 + G_1 \mu_1(\Omega)\right], \label{eqIomm}
\end{align}
with
\begin{equation}
\mu_{i}(\Omega) = \frac{\beta}{\xi_{i}}=\frac{G_1G_2\,\chi_{LC}(\Omega)-\lambda}{\chi_{i}(\Omega)^{-1}-G_{i}^2\,\chi_{LC}(\Omega)}.
\end{equation}
It is evident from Eq. (\ref{spegia2}) that the transduction of voltage input signals into the optical output signal is mainly
determined by the quantity $I(\Omega)$ of Eq.~(\ref{eqIomm}), which is the sum of the two mechanical resonator contributions, i.e., the result of the interference between the two excitation pathways associated with each mechanical mode of Fig.~\ref{fig:1}(a). The quantity $|I(\Omega)|$ determines the voltage sensitivity of the transducer, and larger $|I(\Omega)|$ means higher sensitivity of our transducer, and therefore one has to engineer the couplings $G_j$ in order to realise constructive interference
between the transduction of the two modes and maximise $|I(\Omega)|$.

The explicit expression of the detected spectrum at the output of the transducer simplifies considerably in the following limit: i) $\lambda =0$ (which we have verified is actually satisfied by our experimental setup with a very good approximation); ii) we stop at first order in $G_i$, i.e., we neglect second order terms in $G_i$. In this limit $\chi_{i}^{\rm eff}(\Omega) \to \chi_{i}(\Omega)$, $\mu_i(\Omega) = 0$ and one has the following much simpler output spectrum, and a simpler form of the interference function $I(\Omega)$ in particular,
\begin{align}\label{spegia2b}
	 S_{\rm out}(\Omega)  &=
		\left|\alpha_1 \right|^2 \left | \chi_{1}(\Omega)\right|^2 \,S_{F1}(\Omega)\nonumber \\
		&	+\left|\alpha_2 \right|^2 \left | \chi_{2}(\Omega)\right|^2 \,S_{F2}(\Omega) +S_{\rm in}(\Omega) \\
		& +\left|\alpha_1 G_1 \chi_{1}(\Omega)+ \alpha_2 G_2\chi_{2}(\Omega)\right|^2 \left|\chi_{LC}(\Omega)\right|^2 S_{\delta V}(\Omega). \nonumber
\end{align}
The amplitudes and the relative signs of the couplings $G_{1}$, $\alpha_{1}$, $G_{2}$, and $ \alpha_{2}$ determine the output spectrum and therefore the behavior of the transducer itself. In fact, since the two effective mechanical susceptibilities $\chi_{j}(\Omega)$ halfway between the two mechanical resonance peaks are real and
with opposite signs, we see from Eq.~\eqref{spegia2b} that the products $\alpha_1 G_1$ and $\alpha_2 G_2$ must have the same sign in the case of destructive interference,
and opposite signs in the case of constructive interference between the two transduction pathways.
In the first case we would observe a spectrum region where the RF signal is canceled out by the destructive interference between the transduction of the two mechanical modes. In the second case we would observe a spectrum region between the two resonance peaks where the output signal is
\new{flat, and}
enhanced by the constructive interference between the two electromechanical couplings. Experimentally we observe both behaviors and we refer to Sec.~\ref{sec:Experiment} for further details and discussion.

\subsection{Relation between sensitivity and bandwidth} \label{sensitivity}
The transducer voltage sensitivity can be quantified by appropriately rescaling the detected noise spectrum of Eqs.~(\ref{spegia2}) and (\ref{spegia2b}), i.e., by defining the spectral voltage sensitivity as \cite{Bagci2013}
\begin{equation} \label{eqsens}
	\sqrt{S_{\delta V}^{\rm out}(\Omega)}=  \frac{\sqrt{S_{\rm out}(\Omega)}}{|I(\Omega)| \left|\chi_{LC}(\Omega)\right|}\,.
\end{equation}
One expects that the better the sensitivity the narrower the corresponding bandwidth; we confirm that this is, in fact, the case,
\new{by quantifying this trade-off with a simple formula valid for the constructive interference case. Eq.~\eqref{eqsens} shows that the minimum detectable voltage signal corresponds to the situation of minimum output noise $S_{\rm out}(\Omega)$, and maximum value of the product $|I(\Omega)| \left|\chi_{LC}(\Omega)\right|$. Minimum output noise is achieved when the contribution of all technical noises, that is, thermal and RF ones, are negligible with respect to the unavoidable shot noise contribution, i.e., when the first, second and fourth term in Eq.~\eqref{spegia2b} are negligible with respect to the third term, so that $S_\mathrm{out}(\Omega) \simeq S_\mathrm{in}(\Omega)$. The denominator of Eq.~\eqref{eqsens} is instead maximum when the LC circuit resonance peak (which is typically much broader than the mechanical peaks) is centered between the mechanical doublet, and when $|I(\Omega)|$ is largest, showing why constructive interference is needed for a sensitive transducer. Actually, $|I(\Omega)|$ is maximum exactly at the two mechanical resonance frequencies, where in principle one could get the best sensitivity. However, in order to get a physically meaningful and practical estimation of the optimal detectable voltage, we make here a conservative choice and consider the flat response region obtained by constructive interference between the two mechanical resonances. In fact, the latter resonances do not represent a convenient working point because they are very narrow and extremely sensitive to small frequency shifts, and one expects a quite unstable transducer response there (see also the experimental results in the next Section). Therefore we take as optimal detectable signal the expression of Eq. (\ref{eqsens}) when $S_\mathrm{out}(\Omega) \simeq S_\mathrm{in}(\Omega)$, evaluated halfway between the two mechanical resonances, at $\bar\Omega = (\omega_1 + \omega_2)/2$,
 \begin{equation}\label{mindetsens}
	\sqrt{S^\mathrm{opt}_{\delta V}(\bar\Omega)} =
		\frac{\sqrt{S_\mathrm{in}(\bar\Omega)}}{|I(\bar\Omega)|\chi_\mathrm{LC}(\bar\Omega)|}.
\end{equation}
In the simple case of a symmetric electromechanical system, that is, by assuming same masses ($m = m_1= m_2$), electromechanical couplings ($G = |G_1| = |G_2|$), and damping rates ($\Gamma = \Gamma_1 = \Gamma_2$) for the two mechanical modes, and assuming also optimal optical detection ($\alpha = \alpha_1=\alpha_2=1$), $I(\bar\Omega)$ is given by
 \begin{equation}
	|I(\bar\Omega)| = \frac{G}{m\,\bar\Omega}
		\left| \frac{1}{\rm{i}\Gamma+\Delta\Omega} - \frac{1}{\rm{i}\Gamma-\Delta\Omega} \right|\,,
\end{equation}
where $\Delta\Omega = \omega_2 - \omega_1$. In typical situations one has $\Delta\Omega \gg \Gamma$, so that one can safely write
 \begin{equation} \label{iapp}
	|I(\bar\Omega)|  = \left( \frac{2 G}{m\,\bar\Omega\,\Delta\Omega}\right)\,,
\end{equation}
and replacing this latter expression into Eq.~\eqref{mindetsens}, one finally gets the desired sensitivity--bandwidth--ratio limit
 \begin{equation}\label{s-b-limit}
	\frac{\sqrt{S^\mathrm{opt}_{\delta V}}}{\Delta\Omega} =
	\frac{m\,\bar\Omega\sqrt{S_{\rm in}(\bar\Omega)}}{2|G\,\chi_\mathrm{LC}(\bar\Omega)|} \,.
\end{equation}
This relation shows that, as expected, there is a trade-off between the voltage sensitivity and the bandwidth for an opto-electro-mechanical transducer with a given set of parameters. We can also see that, for a given shot noise level and fixing a desired voltage sensitivity $S^\mathrm{opt}_{\delta V}$, one can increase the bandwidth either by decreasing the mechanical resonator mass, or by increasing the electromechanical coupling, always keeping the LC circuit at resonance so that $|\chi_{\rm LC}|$ is maximum.}

\section{Experiment}\label{sec:Experiment}

\begin{figure}[h!]
\includegraphics[width=.475\textwidth]{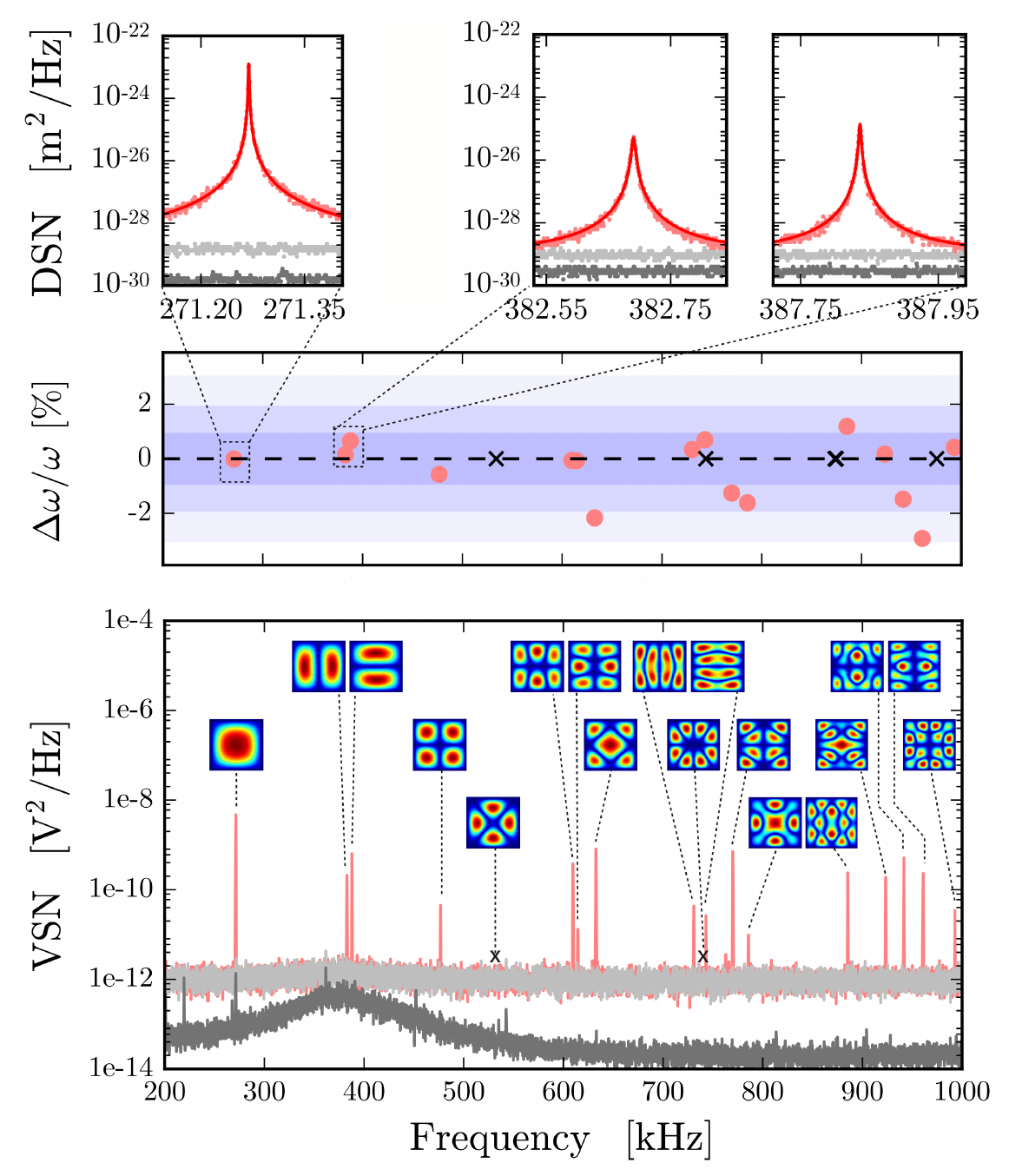}
\caption{
	Top: Calibrated displacement spectral noise (DSN) obtained from the homodyne measurement of the optical output when the system is driven by thermal noise only, i.e., without any electromechanical coupling.
	Top--left, DSN of the first mode (light--red dots) and the theoretical curve (red line) obtained with the best--fit values $\omega_{\mathrm{m}}^{(1,1)} = 2\pi\times\SI{271.269}{\kilo\hertz}$, $\Gamma^{(1,1)}= 2\pi\times\SI{0.9}{\hertz}$, and $\mathrm{m_{opt}}^{(1,1)} = \SI{70.0\pm 0.2}{\nano\gram}$.
	Top--right, the mode doublet exploited for the transduction, with best--fit values  $\omega_{\mathrm{m}}^{(1,2)} = 2\pi\times\SI{382.69}{\kilo\hertz}$, $\Gamma^{(1,2)}=2\pi\times\SI{4.9}{\hertz}$, $\mathrm{m_{opt}^{(1,2)}} = \SI{1.73 \pm .01}{\micro\gram}$, and  $\omega_{\mathrm{m}}^{(2,1)} = 2\pi\times\SI{387.836}{\kilo\hertz}$, $\Gamma^{(2,1)}=2\pi\times\SI{2.6}{\hertz}$, $\mathrm{m_{opt}^{(2,1)}} = \SI{1.18 \pm 0.01}{\micro\gram}$ (see Appendix A).
	Middle: Relative error between the detected frequencies (obtained from the peaks within the broader homodyne spectrum shown at the bottom of the figure) and those obtained from a numerical finite element analysis of the vibrational modes of the metalized membrane. The relative error found for the first five detected modes is less than $1\%$, and for the higher modes less than $3\%$. The black crosses indicate modes which are uncoupled to the light beam and therefore unobservable.
	Bottom: Broad homodyne spectrum. Each peak is associated with the corresponding vibrational mode shape obtained from the finite element analysis.
\new{Light--grey and dark--grey curves denote shot and electronic noise contributions, respectively}.
}\label{fig:2}
\end{figure}

A schematic description of the experiment is given in Fig.~\ref{fig:1}(b). A laser at $\SI{532}{\nano\meter}$ is split into a $\SI{10}{\milli\watt}$ beam (local oscillator), and a few hundreds $ \mu \mathrm{W} $ one for probing the mechanical oscillator.  The beam reflected by the membrane is superposed to the local oscillator for detecting the phase--fluctuations. The low--frequency region of the voltage spectral noise of the homodyne signal is exploited to lock the interferometer to the
\new{grey fringe (i.e., in the condition where the interferometer output is proportional to the membrane displacement)}
by means of a PID control. The thermal displacement of the metalised membrane modes are revealed in the high--frequency range, as shown in Fig.~\ref{fig:2}. Calibration and fitting of the zoomed spectra of the fundamental and doublet modes shown in Fig.~\ref{fig:2} allows us to obtain the
\new{optical masses $\mathrm{m_{opt}^{(1,1)}}$, $\mathrm{m_{opt}^{(1,2)}}$ and $\mathrm{m_{opt}^{(2,1)}}$ of each mode. As explained in Appendix A, they are given by $\mathrm{m_{opt}^{(i,j)}}=\mathrm{m_{eff}}/\mathrm{\alpha_{(i,j)}}^2$, that is, by the effective mass of these three membrane modes, which are all equal, divided by the square of the respective optomechanical coupling. One can estimate from them the most likely value of the center of the laser beam (and therefore of the optomechanical couplings $\mathrm{\alpha_{(i,j)}}$), and of the effective mass. The latter is equal to $\mathrm{m_{eff}} \simeq \SI{67.3}{\nano\gram}$, in very good agreement with the prediction of finite element method (FEM) numerical analysis of the metalised membrane.}

The membrane is placed on top of a four--segment copper electrode to form a variable capacitor $C_\mathrm{m}(\{x_i\})$, which depends  upon the transversal displacement of the membrane and therefore on the two mode displacements, $x_i$.
The distance $h_0$ between the metalized membrane and the four--segment electrodes has been determined to be $\SI{31.0\pm.1}{\micro\meter}$ by the measurement of the frequency shift of the membrane fundamental mode $(1,1)$ as a function of the applied $V_{\rm DC}$, and the estimation of the effective area of the membrane capacitor (see Appendix B and~\cite{Bagci2013}).
This capacitor is added in parallel to the rest of capacitors of the LC circuit $C_{0}$, and the total capacitance, $C(\{x_i\})=C_{0}+C_\mathrm{m}(\{x_i\})$, is connected in parallel to a ferrite core inductor with inductance of $ L\simeq\SI{427}{\micro\henry}$. Taking into account the total series resistance of contacts and wires $R$, we have an LC resonator with resonance frequency $\Omega_\mathrm{LC}/2\pi \simeq \SI{383}{\kilo\hertz}$, therefore quasi--resonant with the two mechanical modes, and a quality factor $Q \simeq 81$.

\begin{figure}[h!]
\includegraphics[width=.475\textwidth]{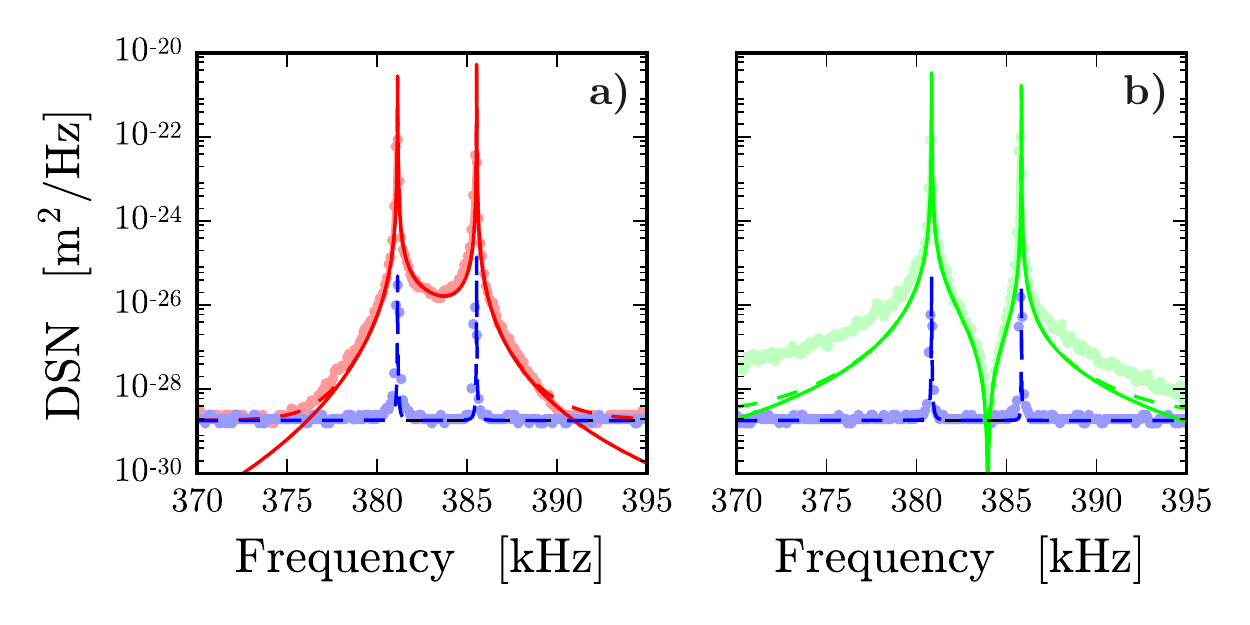}
\caption{Displacement spectral noise (DSN). Light--red and light--green dots correspond to the detection of constructive (a) and destructive (b) interference between the two mechanical transduction pathways, respectively, in the presence of an applied voltage bias $V_{\rm DC}=\SI{270}{\volt}$ and an input RF signal.
Light--blue dots refer to the optical output spectrum due to thermal noise and
\new{without any RF input to the LC circuit}.
Solid  red and green lines are the theoretical expectations without noise.
Dashed red, green, and blue lines account also for the shot--noise contribution.}
	\label{fig:3}
 \end{figure}

We have studied the behaviour of our device as a high--sensitive optical detector of RF--signals by fixing the applied dc--bias at $V_{\rm DC}=\SI{270}{V}$. A broadband RF signal was injected into the system inductively using an auxiliary inductor in front of the main LC inductor, and the corresponding displacement spectral noise (DSN) was detected. The nonzero voltage bias couples the LC circuit with the two mechanical modes whose motion, in turn, modulates the phase of the light; as a result, the input RF signal is transduced as an optical phase modulation readout by the interferometer. The results are shown in Fig.~\ref{fig:3}, where the two plots correspond to two different electrode configurations.

The comparison evidences that by changing the electrodes on which the bias voltage is applied we are able to control the interference between the two transduction pathways associated with each mechanical mode schematically illustrated in Fig.~\ref{fig:1}. In particular, by changing the electrodes, we are able to change the effective areas of the membrane capacitor, thereby changing the relative sign between the two electromechanical couplings $G_1$ and $G_2$ (see Appendix B). As discussed in Sec.~\ref{theory}, this relative sign flip corresponds to switching from constructive interference (Fig.~\ref{fig:3}a) to destructive interference (Fig.~\ref{fig:3}b). In fact, in our case the two optomechanical couplings $\alpha_i$ are positive numbers (see below), and $G_1$ and $G_2$ have the same sign in the case of destructive interference (Fig.~\ref{fig:3}b) and opposite signs in the case of constructive interference (Fig.~\ref{fig:3}a). This fact is confirmed by the theoretical curves which best overlap with the experimental data, corresponding to the following values of the electromechanical couplings:
$G_1=\SI{118.41}{\volt\per\meter}$ and $G_2=\SI{-115.31}{\volt\per\meter}$ for the red--line in Fig.~\ref{fig:3}a, and $G_1=\SI{117.63}{\volt\per\meter}$ and $G_2=\SI{110.89}{\volt\per\meter}$ for the green--line in Fig.~\ref{fig:3}b. These values have been confirmed within an $8\%$ error, with an independent method based on the explicit evaluation of the membrane--electrode capacitance and its derivatives from the knowledge of the device geometry (see Appendix B and Ref.~\cite{Bagci2013}). This geometrical estimation of the electromechanical coupling crucially depends upon the overlap between the electrodes and the positive and negative portions of the chosen membrane vibrational eigenmode, and therefore also provides an idea of how one can control the relative sign between the two electromechanical couplings by applying the voltage bias to different electrodes.

The position of the laser beam with respect to the membrane determines the transverse overlap between the optical laser field and each mechanical mode, and therefore the optomechanical couplings $\alpha_i$ giving the weight of the two interference pathways. As shown in Appendix A, we have found for the constructive interference case the best values $\alpha_1 = \num{0.196}$ and $\alpha_2 = \num{0.240}$, while for the destructive interference we found $\alpha_1 = \num{0.196}$ and $\alpha_2 = \num{0.121}$. The theoretical prediction is less accurate away from the mechanical resonances for the destructive case; in this latter case in fact one has constructive interference effects between the doublet modes and the fundamental and higher mechanical modes, which are not fully taken into account by our model.

We remark that the possibility to tune the performance of our two--mode transducer by controlling the relative sign of the electromechanical couplings and the associated interference effect is available only when the two mechanical modes are simultaneously coupled to \emph{two distinct} electromagnetic modes. In fact, if we would have simplified the scheme and used a \emph{unique} electromagnetic (either radio--frequency or optical) mode both for coupling the modes and reading out the signals, we would always get a destructive interference pattern in the output spectrum, and the constructive case of Fig.~\ref{fig:3}a would be impossible. In such a case $\alpha_i$ and $G_i$ ($i=1,2$) share the same sign, and therefore the response of the two mechanical modes in the frequency band within the two resonances would always be out of phase~\cite{Caniard2007,Cerdonio2001}.
Our two--mode transducer has analogies with the devices recently proposed in Refs.~\cite{Xu2016,Bernier2016,Peano2015} for nonreciprocal conversion between microwave and optical photons, and demonstrated in Refs.~\cite{Bernier2016,Peterson2017}, in which two mechanical modes are simultaneously coupled via four appropriate drives with two different microwave cavity modes, for nonreciprocal signal conversion between the latter.
In our case the configuration corresponding to the constructive interference of Fig.~\ref{fig:3}a realises the unidirectional transduction of RF signals into optical ones, while the one corresponding to the destructive interference of Fig. 3b realises an isolator which, within the bandwidth where $I(\Omega) \simeq 0 $, inhibits the transmission of RF signals to the optical output. The device demonstrated here has the advantage that it does not require driving with four different tones and the validity of the rotating wave approximation. Moreover, the device is easily reconfigurable because one can switch from one configuration to the other by simply switching electrodes~\cite{OurNote}.
 \begin{figure}[h!]
\includegraphics[width=.325\textwidth]{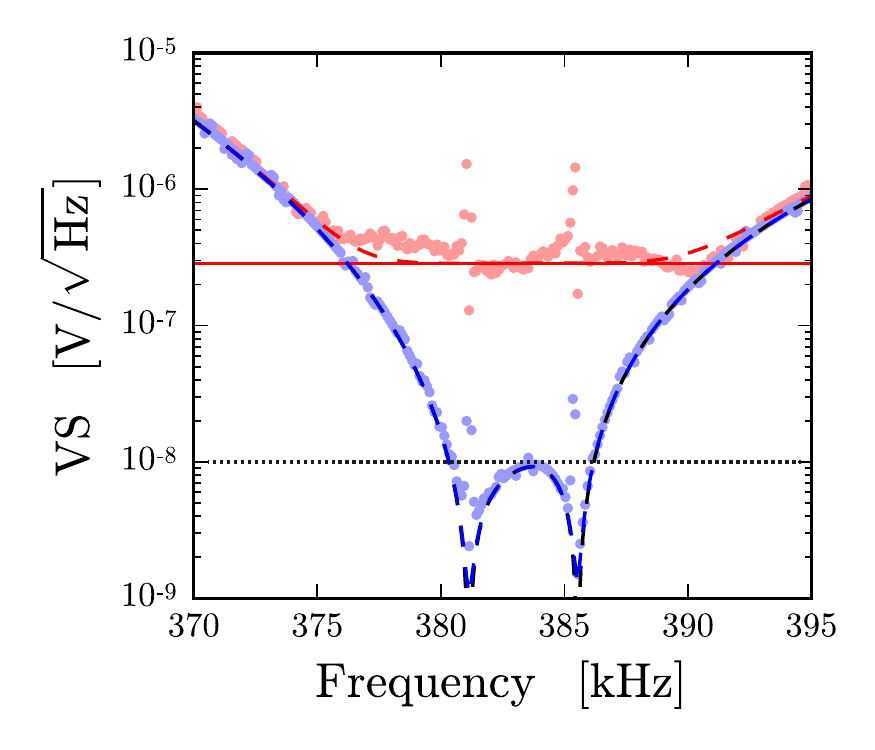}
\caption{Voltage sensitivity (VS) of the RF--to--optical transducer.
  Light--red dots correspond to the inferred voltage sensitivity of our transducer from the blue data of Fig.~\ref{fig:3}a, that is the square--root of the DSN divided by the interface response function, which is equal to $\SI{300}{\nano\volt\per\sqrt{\hertz}}$ over a bandwidth of $\SI{15}{\kilo\hertz}$. The light--blue dots represent the
  \new{optimal sensitivity achieved by our device in the case of negligible RF noise, equal to $\SI{10}{\nano\volt\per\sqrt{Hz}}$ over a bandwidth of 5 kHz, dotted--black line. Dashed and solid lines represent the corresponding theoretical expectations as in Fig.~\ref{fig:3}. In this latter case the sensitivity--bandwidth ratio is in agreement with the optimal limit given by Eq. (\protect\ref{s-b-limit})}.
  }\label{fig:4}
 \end{figure}

As we have discussed in Sec.~\ref{theory}, under the condition of constructive interference the mechanical modes are responsible for an improved transduction of RF signals into the optical output within the frequency band between the two mechanical resonances. Therefore we expect that under the conditions of Fig.~\ref{fig:3}a the device acts as a transducer with an increased bandwidth. This is confirmed by Fig.~\ref{fig:4} where we show the voltage sensitivity (VS) defined in Eq.~(\ref{eqsens}), i.e., the minimum detectable voltage, corresponding to the total noise spectrum of Fig.~\ref{fig:3}a divided by the interface response function.
The red--light circles corresponds to the broader band voltage sensitivity of our transducer which is equal to $\SI{300}{\nano\volt\per\sqrt{\hertz}}$ over a bandwidth of 15 kHz between the two modes, obtained in the case when RF noise dominates over thermal and shot noise. The blue dots and lines instead correspond to the
\new{optimal}
sensitivity of our device, around $\SI{10}{\nano\volt\per\sqrt{Hz}}$ over a bandwidth of 5 kHz,
\new{achieved}
in the
\new{opposite limit when the contribution of input RF noise is negligible with respect to thermal and shot noise. In this latter limit, in the flat region between the two resonance peaks thermal noise is also negligible, and the data (blue dots) exactly satisfy the optimal sensitivity--bandwidth ratio of Eq.~\eqref{s-b-limit}. Instead, the data (red dots) in the presence of a non--negligible noise contribution from the LC circuit, corresponds to a larger value of the sensitivity--bandwidth ratio compared to the optimal value. For example in Fig.~\ref{fig:4}, the red data correspond to a sensitivity--bandwidth ratio 10 times larger than the optimal one achieved by the blue data.}

\begin{figure}[h!]
\includegraphics[width=.325\textwidth]{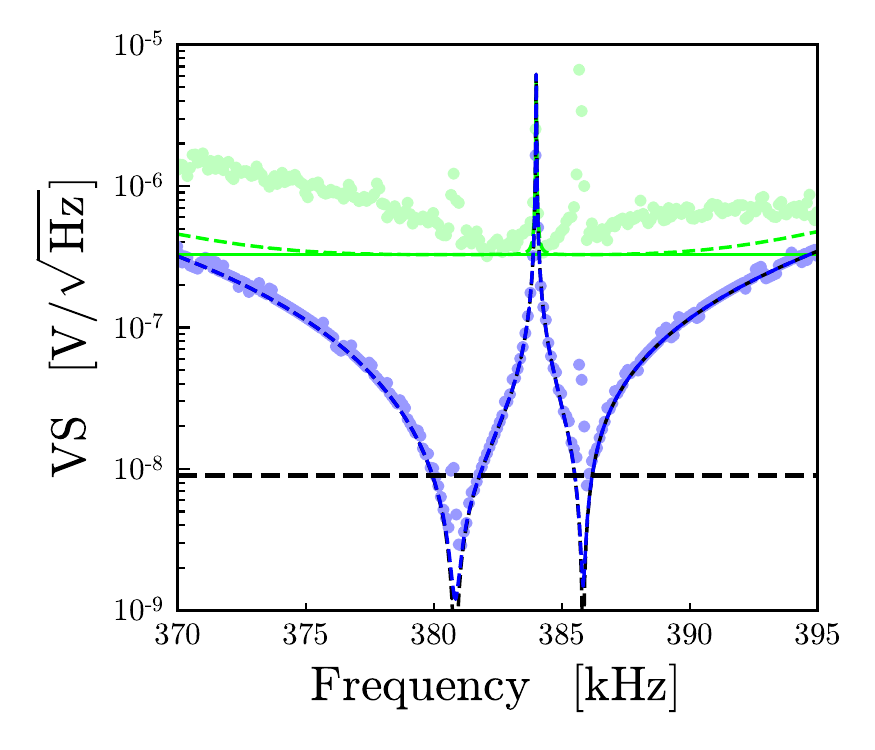}
\caption{
\new{Voltage sensitivity (VS) of the RF--to--optical transducer in the presence of destructive interference, from the data of Fig. \protect\ref{fig:3}b.
  Light--green dots correspond to the inferred voltage sensitivity of our transducer, that is the square--root of the DSN divided by the interface response function, which is equal to $\SI{300}{\nano\volt\per\sqrt{\hertz}}$ over a bandwidth approximately equal to $\SI{5}{\kilo\hertz}$. The light--blue dots represent the sensitivity achieved in the case of negligible RF noise, which tends to diverge at the frequencies where one has destructive interference and the device is not sensitive to the input RF signal. Dashed and solid lines represent the corresponding theoretical expectations}.
  }\label{fig:5new}
 \end{figure}

\new{For comparison, in Fig.~\ref{fig:5new} we show the minimum detectable voltage in the \emph{destructive} interference case of Fig.~\ref{fig:3}b. Even though in the case of large RF noise we have a comparable sensitivity of $\sim \SI{300}{\nano\volt\per\sqrt{\hertz}}$ to that of the constructive interference case, the situation is completely different in the regime of negligible RF input noise (blue dots and theoretical curve). As expected, in this latter case, the sensitivity significantly worsens between the two mechanical resonances and the minimum detectable voltage tends to diverge in correspondence to the destructive interference condition where the device acts as an isolator with respect to the RF input. It is evident that in the presence of destructive interference the device cannot be operated as an RF--to--optical transducer and that a sensitivity--bandwidth ratio cannot be even defined here.}

\new{We also remark that the present transducer can also be treated as a radiofrequency amplifier, transforming a voltage input signal into a voltage signal at the output of the optical detector, but at much higher signal to noise ratio,  with a given gain and a given input impedance. At the working point described here, and corresponding to Fig.~\ref{fig:4} and~\ref{fig:5new}, we have measured for our device a gain of $\SI{30}{db}$ at the mechanical frequencies, and a gain of $\SI{10}{db}$ in the frequency range between them.  Moreover we have characterized the input impedance obtaining a value $Z_{\rm in}=\SI{51.2 + i19.5}{\kilo\ohm}$}.

\subsection{Improving the two-mode transducer performance} \label{performance}
\begin{figure}[h!]
\includegraphics[width=0.475\textwidth]{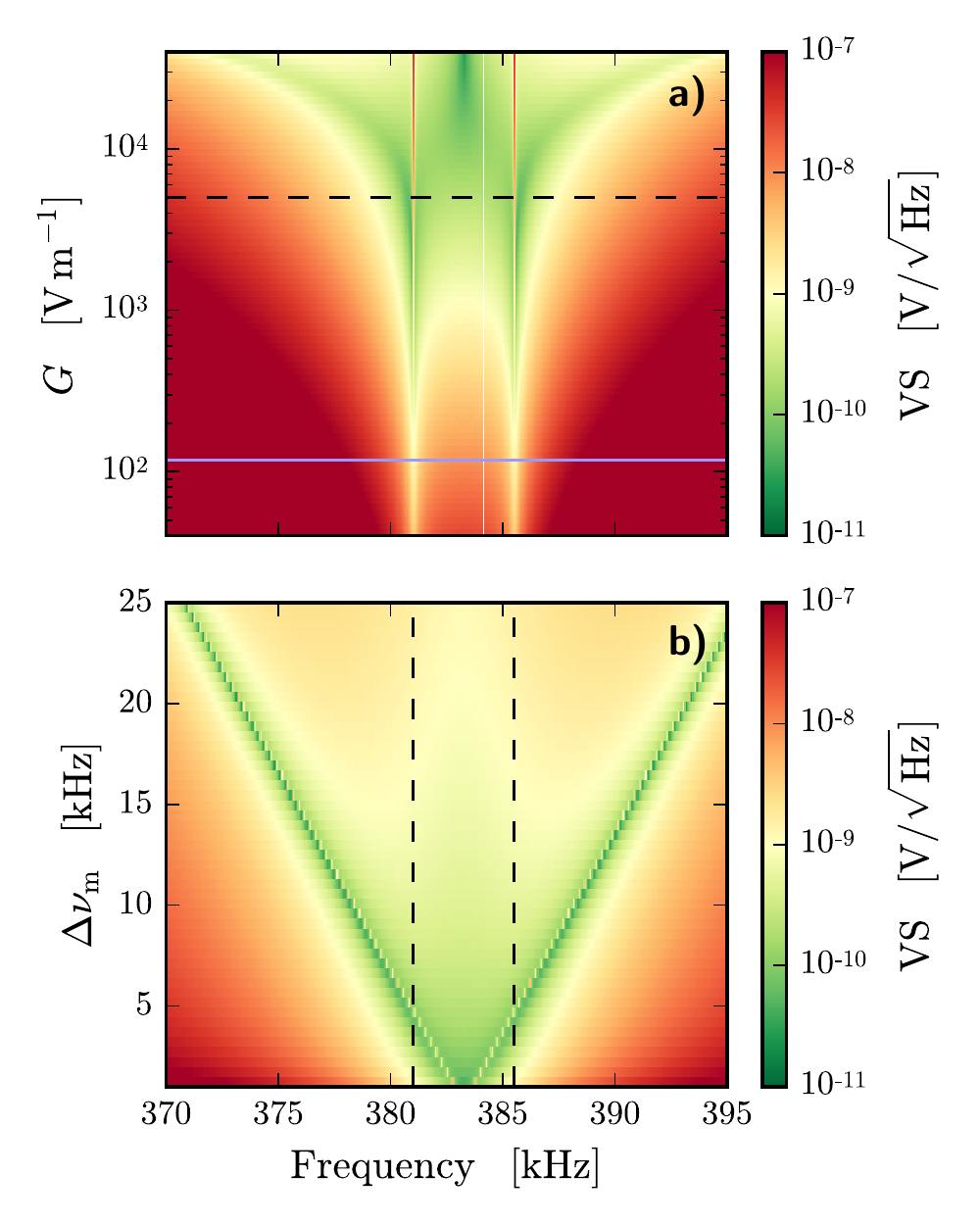}
\caption{
	Theoretical prediction for the voltage sensitivity (VS) of an RF--to--optical transducer based on a two-mode mechanical resonator in the case of negligible RF noise with a shot noise level $S_{\mathrm{in}}=1.8\times10^{-29}\, \rm m^2/Hz$.
	Top:  VS as a function of the frequency and of the electromechanical coupling $G$ (assumed to be equal in modulus for the two modes). The other parameters have been chosen to be very close to those of our experimental device. The two vertically brighter features represent the mechanical mode resonance frequencies at $\omega_\mathrm{m}^{(1,2)} = 2\pi \times \SI{381}{\kilo\hertz}$ and $\omega_\mathrm{m}^{(2,1)} = 2\pi \times \SI{385.5}{\kilo\hertz}$, with same damping rate $ \Gamma_\mathrm{m}^{(1,2)}= \Gamma_\mathrm{m}^{(2,1)}  = 2\pi\times\SI{3.6}{\hertz}$, same effective mass $\mathrm{m}_\mathrm{eff} = \SI{67.3}{\nano\gram}$, and same optomechanical coupling $\alpha_1 = \alpha_2 = \num{.194}$. The LC circuit resonates at $\omega_\mathrm{LC} = [\omega_\mathrm{m}^{(1,2)} + \omega_\mathrm{m}^{(2,1)}]/2$ with a quality factor
$Q = \num{81.5}$. The blue solid line denotes the electromechanical coupling in our device, $G = \SI{118}{\volt\per\meter}$, while the black dashed line denotes the value $G = \SI{5}{\kilo\volt\per\meter}$ needed to obtain a mean voltage sensitivity of the order of 1 $\si{\nano\volt\per\sqrt{\hertz}}$ over a bandwidth of $\SI{15}{\kilo\hertz}$. For larger $G$ both the sensitivity and the bandwidth increase.
	Bottom: VS as a function of the frequency and of the mechanical mode separation $\Delta\nu_\mathrm{m} = [\omega_\mathrm{m}^{(2,1)} - \omega_\mathrm{m}^{(1,2)}]/2\pi$ evaluated for the same parameters as the plot above, and with a value of the electromechanical coupling $G$ indicated by the black dashed line of the top figure. The vertical black dashed lines represent the two resonance frequencies chosen above. We see that one can achieve and maintain a voltage sensitivity of around 1 $\si{\nano\volt\per\sqrt{\hertz}}$ over a bandwidth which increases for increasing frequency separation between the two mechanical modes.}
	\label{fig:0}
 \end{figure}

Using the theoretical description provided in Sec.~\ref{theory} we now see how much one could improve the performance of our transducer in the constructive interference configuration. The voltage sensitivities achievable in a device similar to that experimentally demonstrated here, but with tunable electromechanical couplings $|G_1|=|G_2|=G$ and frequency separation $\Delta\nu_\mathrm{m} = [\omega_\mathrm{m}^{(2,1)} - \omega_\mathrm{m}^{(1,2)}]/2\pi$ are shown in Fig.~\ref{fig:0}.
We show the transducer voltage sensitivity as a function of the electromagnetic coupling $G$ at a fixed mechanical mode frequency separation $\Delta\nu_\mathrm{m}$ in Fig.~\ref{fig:0}(a), and versus the mechanical mode splitting at a fixed $G$ in Fig.~\ref{fig:0}(b) in the case of negligible RF noise.  The voltage sensitivity is calculated from Eqs.~(\ref{spegia2}) and (\ref{eqsens}) considering the following experimental parameters: equal effective mass $\mathrm{m}_{\mathrm{eff}}=\SI{67.3}{\nano g}$, equal mechanical damping rates $\Gamma=2\pi \times 3.6$ Hz, equal optomechanical couplings $\alpha_{1}=\alpha_{2}=0.194$, an LC circuit resonating halfway between the two mechanical resonances with a quality factor
$Q = \num{81.5}$, and a shot noise level $S_{\mathrm{in}}=1.8\times10^{-29}\, \rm m^2/Hz$. In Fig.~\ref{fig:0}(a) the two resonance frequencies are fixed at $\omega_\mathrm{m}^{(1,2)}/2\pi = \SI{381}{kHz}$ and $\omega_\mathrm{m}^{(2,1)}/2\pi=\SI{385.5}{ kHz}$. The blue-solid line denotes the electromechanical coupling in our device, $G=\SI{118}{V m^{-1}}$, while the dashed-line denotes the electromechanical coupling $G=5$ kV m$^{-1}$, which is used to calculate  Fig.~\ref{fig:0}(b).
Fig.~\ref{fig:0}(a) shows that, as expected, both sensitivity and bandwidth can be increased by increasing the electromechanical coupling and that one can achieve sensitivities comparable to those of Ref.~\cite{Bagci2013} over a larger bandwidth in the strong coupling regime where the LC and the mechanical modes hybridize, which occurs in our case when $G > 10$ kV m$^{-1}$. A feasible way to achieve these values of the coupling is to decrease the distance $d$  between the electrodes and the metalized SiN membrane since the coupling scales as the inverse square of $d$, and this strong coupling regime could be achieved with a distance $d \simeq \SI{3}{\micro m}$.
Fig.~\ref{fig:0}(b) instead shows that even in a regime away from the strong coupling regime, the transduction bandwidth can be increased simply by increasing the mechanical mode frequency splitting. With a coupling $G=5$ kV m$^{-1}$, about a factor of 30 larger than the one showed by our device, a sensitivity of the order of 1 $\si{\nano\volt\per\sqrt{\hertz}}$  is reachable over a bandwidth that depends essentially only upon the mechanical mode splitting. In practice, by improving the device demonstrated here, for example by operating at a membrane capacitor distance of around  $d \simeq \SI{3}{\micro m}$  in order to reach $G \simeq 10$ kV m$^{-1}$, and increasing the mechanical mode frequency splitting by using a rectangular membrane of $0.9\times 1.1 \,\rm mm$ sides, one could achieve the same sensitivity of 800 $\si{\pico\volt\per\sqrt{\hertz}}$ of Ref.~\cite{Bagci2013} over a larger bandwidth of $40$ kHz.

\section{Conclusions} \label{Concl}

We have theoretically shown and experimentally demonstrated that one can engineer a constructive interference between two or more mechanical modes coupled to the same resonant LC circuit in order to increase the transduction bandwidth of an RF--to--optical transducer with a target voltage sensitivity equal to that of the single mechanical mode transducer. We have presented here a proof--of--principle experiment with a first generation device proving the reliability of the proposed technique and its physical insight. We have seen that an improved version of the same device could outperform existing single--mode opto--electro--mechanical transducer in terms of sensitivity and especially in terms of bandwidth. The proposed multimode transducer based on constructive interference is advantageous and more flexible with respect to the one based on a single mechanical mode. In fact, in single--mode opto--electro--mechanical transducers bandwidth and sensitivity are strongly related and determined only by the electromechanical coupling. In the case of capacitive coupling, it is extremely hard to achieve very large values of such a coupling because the bias voltage and the membrane capacitor area cannot be too large, and it is hard to reach membrane capacitor distances well below one micron. On the contrary, in multimode opto--electro--mechanical transducers in the constructive interference configuration, for a given voltage sensitivity, the bandwidth is mainly determined by the mechanical frequency splitting and therefore it can be significantly increased even without entering the strong electromechanical coupling regime.

\begin{acknowledgments}
We acknowledge the support of the European Commission through the ITN--Marie Curie project cQOM, the FP7 FET-Open Project n. 323924 iQUOEMS, and the H2020--FETPROACT--2016 project n. 732894 ``HOT''. We also thanks Norcada for providing us the Nb metalized SiN membranes.
\end{acknowledgments}

\appendix

\section{Data analysis}
\subsection{Determination of mechanical parameters from thermal noise spectra}

\begin{figure}[h]
\centering
\includegraphics[width=0.35\textwidth]{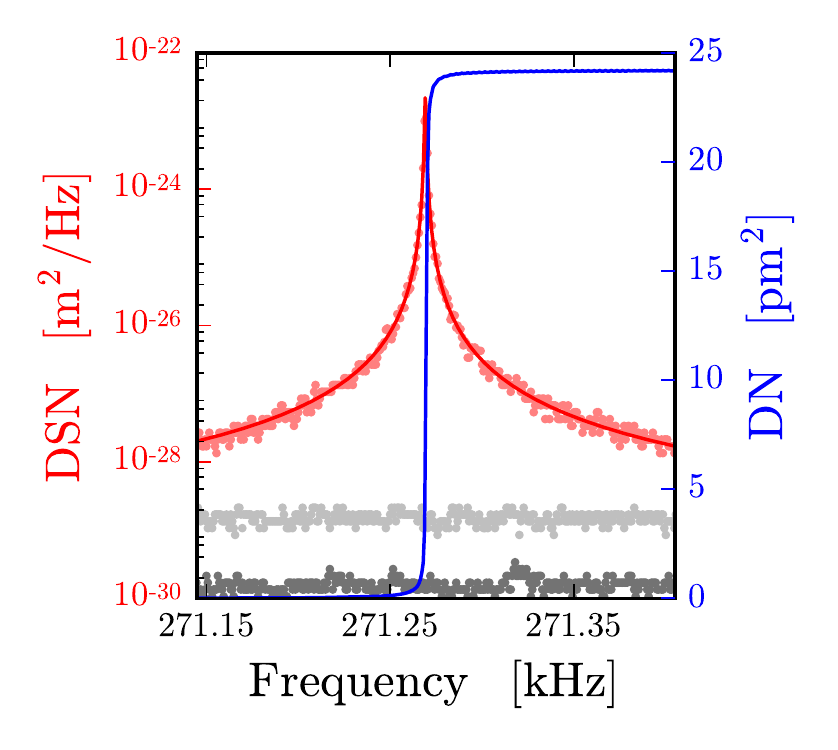}
\caption{Displacement spectral noise (DSN) for the fundamental mode $(1,1)$. The calibration parameter is $V_{pp}= \SI{2.7}{\volt}$, and the best--fit values are $\omega_{\mathrm{m}}^{(1,1)} = 2\pi\times\SI{271.269}{\kilo\hertz}$, $\Gamma^{(1,1)}= 2\pi\times\SI{0.9}{\hertz}$, and $\mathrm{m^{(1,1)}_{opt}} = \SI{70.0\pm 0.2}{\nano\gram}$. The size of the step in the displacement noise (DN) (blue curve), that is the marginal of the DSN, determines the variance of the mechanical displacement $\langle x^2 \rangle^{(1,1)}$ to be $\SI{24.18}{\square\pico\meter}$.}
\label{fig:Spectra11}
\end{figure}

\begin{figure}[h]
\centering
\includegraphics[width=0.5\textwidth]{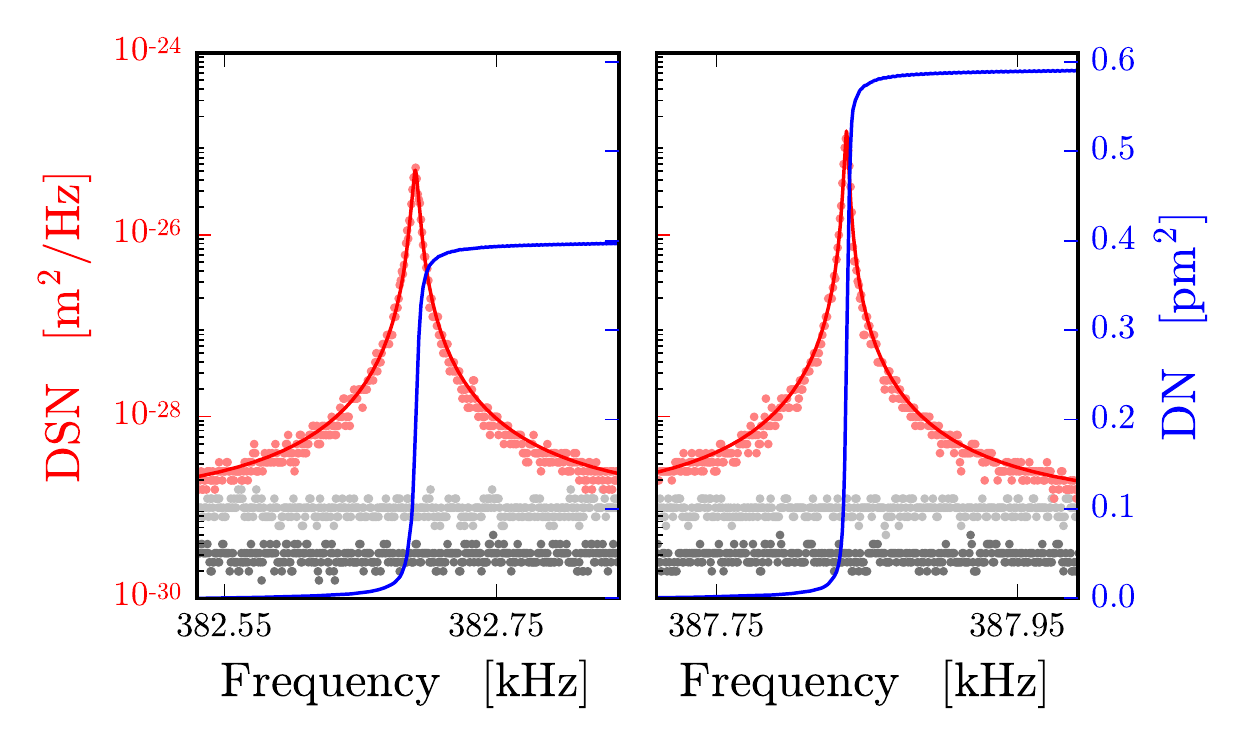}
\caption{Displacement spectral noise (DSN) for the doublet $(1,2)-(2,1)$. The calibration parameter is $V_{pp}= \SI{2.8}{\volt}$. The best--fit values are $\omega_{\mathrm{m}}^{(1,2)} = 2\pi\times\SI{382.69}{\kilo\hertz}$, $\Gamma^{(1,2)}=2\pi\times\SI{4.9}{\hertz}$, $\mathrm{m_{opt}^{(1,2)}} = \SI{1.73 \pm .01}{\micro\gram}$, and  $\omega_{\mathrm{m}}^{(2,1)} = 2\pi\times\SI{387.836}{\kilo\hertz}$, $\Gamma^{(2,1)}=2\pi\times\SI{2.6}{\hertz}$, $\mathrm{m_{opt}^{(2,1)}} = \SI{1.18 \pm 0.01}{\micro\gram}$. The variances of the mechanical displacement are $\langle x^2 \rangle^{(1,2)} \simeq \SI{0.397}{\square\pico\meter}$, and $\langle x^2 \rangle^{(2,1)} \simeq \SI{0.590}{\square\pico\meter}$.}
\label{fig:Spectra}
\end{figure}

The mechanical properties of the membrane vibrational modes, that is, their resonance frequency, damping rate and mass, can be extracted from the measured homodyne spectra in the presence of thermal noise only, that is, in the absence of the electromechanical coupling, occurring when $ V_{\rm DC}=0$ and the RF signal is turned off.
For a generic harmonic oscillator of mass $\mathrm{m}$, frequency $\omega_\mathrm{m}$ and damping $\Gamma$ in the presence of thermal noise at temperature $T$, the variance of its mechanical displacement, $ \langle x^2 \rangle = k_BT/\mathrm{m}\,\omega_\mathrm{m}^2$, is related to the displacement spectral noise (DSN) $S_{xx}(\omega)$ by the relation
\begin{equation}
  \langle x^2 \rangle = \int_{-\infty}^{+\infty}S_{xx}(\omega)\frac{d\omega}{2\pi}
  				= \int_{0}^{+\infty}\bar{S}_{xx}(\nu)d\nu\,,
\end{equation}
where
\begin{equation}
	S_{xx}(\omega)=\frac{2\,\mathrm{m}\,\Gamma\,k_BT}
	{|\mathrm{m}(\omega_\mathrm{m}^2 - \omega^2 - \rm{i}\omega\Gamma)|^2}\,,
\end{equation}
and defining $\omega = 2\pi \nu$, $\omega_\mathrm{m} = 2\pi \nu_\mathrm{m}$, and $\Gamma = 2\pi \gamma$, one has
\begin{equation} \label{theospe}
	\bar{S}_{xx}(\nu)=\frac{1}{\pi \mathrm{m}}\frac{2\,\gamma\,k_BT}
	{|\nu_\mathrm{m}^2 - \nu^2 - \rm{i}\nu\gamma|^2}\,\,.
\end{equation}
The \emph{measured} DSN, $\bar{S}_{xx}^{(m)}(\nu)$, is obtained from the calibration of the voltage spectral noise $S_{VV}(\nu)$ effectively detected at the output of our optical interferometer,
\begin{equation}
	\bar{S}_{xx}^{(m)}(\nu)= S_{VV}(\nu)\,G_{xV}^2\,,
\end{equation}
with the calibration factor $G_{xV} = \lambda/(2\pi V_{pp})$, where  $\lambda = \SI{532}{\nano\meter}$ is the laser wavelength used, and $V_{pp}$ is the peak--to--peak voltage value of the interferometer interference fringes. Then the measured DSN is fitted with the theoretical $\bar{S}_{xx}(\nu)$ of Eq.~[\ref{theospe}] obtaining best--fit values for $\omega_{\mathrm{m}}$ and $\Gamma$.
\new{Due to the effect of the optical transduction (see Eq.~(\ref{eq:detected})), for each mechanical mode the fit provides for the mass the value of what can be called the \emph{optical} mass $\mathrm{m^{(n,m)}_{opt}}$, which is related to the physical effective mass of each mode and the optomechanical coupling $\mathrm{\alpha_{(n,m)}}$ by the relation $\mathrm{m^{(n,m)}_{opt}} = \mathrm{m^{(n,m)}_{eff}}/\mathrm{\alpha_{(n,m)}^2}$}.
The variance of the mechanical displacement $\langle x^2 \rangle $ is instead equal to
\new{the size of the step}
in the measured displacement noise (DN), that is, the marginal of the DSN, (see the blue curves in Figs.~\ref{fig:Spectra11}--\ref{fig:Spectra}). We have performed such a fit for the fundamental vibrational mode of the membrane $(1,1)$ (see Fig.~\ref{fig:Spectra11}), and for the first excited vibrational doublet $(1,2)$ and $(2,1)$ exploited here for our transducer (see Fig.~\ref{fig:Spectra}).

For the $(1,1)$ mode we obtained the best--fit values $\omega_{\mathrm{m}}^{(1,1)} = 2\pi\times\SI{271.269}{\kilo\hertz}$, $\Gamma^{(1,1)}= 2\pi\times\SI{0.9}{\hertz}$, and $\mathrm{m_{opt}^{(1,1)}} = \SI{70.0\pm 0.2}{\nano\gram}$. The size of the step in the displacement noise (DN) yielded $\langle x^2 \rangle ^{(1,1)}\simeq \SI{24.18}{\square\pico\meter}$.
For the $(1,2)-(2,1)$ doublet instead we obtained the best--fit values $\omega_{\mathrm{m}}^{(1,2)} = 2\pi\times\SI{382.69}{\kilo\hertz}$, $\Gamma^{(1,2)}=2\pi\times\SI{4.9}{\hertz}$, $\mathrm{m^{(1,2)}_{opt}} = \SI{1.73 \pm .01}{\micro\gram}$, and  $\omega_{\mathrm{m}}^{(2,1)} = 2\pi\times\SI{387.836}{\kilo\hertz}$, $\Gamma^{(2,1)}=2\pi\times\SI{2.6}{\hertz}$, $\mathrm{m_{opt}^{(2,1)}} = \SI{1.18 \pm 0.01}{\micro\gram}$. The variances of the mechanical displacement are $\langle x^2 \rangle^{(1,2)} \simeq \SI{0.397}{\square\pico\meter}$, and $\langle x^2 \rangle^{(2,1)} \simeq \SI{0.590}{\square\pico\meter}$.

\subsection{ Determination of the effective mass and of the optomechanical couplings}
\new{The effective mass $\mathrm{m_{eff}}$ associated with a vibrational mode depends in general upon the mode volume, and in the case of a thin membrane it can be written as}
\begin{equation}\label{realmass}
  \mathrm{m^{(n,m)}_{eff}}=\int\int d\mathrm{x} d\mathrm{y}\,\sigma(x,y)\,u_{\mathrm{(n,m)}}(x,y)^2,
\end{equation}
where $\sigma(x,y)$ is the average mass surface density of the membrane and $u_{\mathrm{(n,m)}}(x,y)$ is the dimensionless eigenfunction of the vibrational mode with indices $\mathrm{(n,m)}$ \cite{Biancofiore2011}.
\new{As discussed in the previous subsection, the masses obtained from the fitted thermal noise spectra instead depend also upon the optomechanical couplings $\alpha_{\mathrm{(n,m)}}$}
which differ from one mode to another because the laser beam illuminates a certain spot on the membrane, where different modes have different displacement
amplitudes. After calibration of the DSN, the couplings $\alpha_{\mathrm{(n,m)}}$ coincide with the dimensionless \emph{transverse overlap} parameters \cite{Biancofiore2011}, given by
\begin{equation}\label{overlapgen}
  \alpha_{\mathrm{(n,m)}}(x,y)=\int_0^L d\mathrm{x'}\int_0^L d\mathrm{y'}u_{\mathrm{(n,m)}}(x',y')I(x,y,x',y'),
\end{equation}
where $I(x,y,x',y')$ is the normalised intensity profile of a laser beam centered at $(x,y)$, and $L$ is the length of the side of the square membrane. In the case of our experiment, the theoretical value of Eq.~\eqref{overlapgen} can be analytically evaluated because we used a $\mathrm{TEM}_{00}$ Gaussian beam with waist $w$ at the membrane position, and we have verified with a finite element method analysis that for the first three vibrational modes studied here the homogeneous membrane eigenmodes, $u_{\mathrm{(n,m)}}(x,y)=\sin (n\pi x/L) \sin (m\pi y/L)$, provide a very good approximation.
Assuming optical losses from clipping negligible, the domain of integration can be extended to the entire plane, and
one gets from Eq.~\eqref{overlapgen}
\begin{equation}
	\alpha_\mathrm{nm}^{\rm{(th)}}(x,y)  = {\rm e}^{-w^2(k_n^2 + k_m^2)/8}\,\sin(k_nx)\,\sin(k_my)\,,
\end{equation}
where $k_{n} = n\,\pi/L$, and $k_m = m\,\pi/L$, which depend upon the unknown beam center $(x,y)$.

One can get a very good estimate of the beam center position $(x,y)$ (and therefore of the transverse overlaps and of the physical effective masses $\mathrm{m^{(n,m)}_{eff}}$) in our setup, by applying a treatment analogous to that of Ref.~\cite{Nielsen2017}. For each of the three detected vibrational modes, the variance of the mechanical displacement $\langle x^2 \rangle^{(n,m)}$ provides an indirect estimate $(\bar x,\bar y)$ of $(x,y)$, because
\begin{equation}
	\langle x^2 \rangle^{(n,m)}
		= \frac{k_BT}{\mathrm{m^{(n,m)}_{opt}}\,\omega^2_\mathrm{nm}}
		=\alpha^2_\mathrm{nm}(\bar x,\bar y)\frac{k_BT}{\mathrm{m^{(n,m)}_{eff}}\,\omega^2_\mathrm{nm}}\,,
\end{equation}
from which one derives the experimental estimate
\begin{equation}\label{alphaesto}
	\alpha_\mathrm{nm}^{\rm{(ex)}}(\bar x,\bar y)
	\simeq \sqrt{\langle x^2 \rangle^{(n,m)}\,\mathrm{m^{(n,m)}_{eff}}\,\omega^2_\mathrm{nm}/k_BT}\,,
\end{equation}
which depends upon the measured quantities $\langle x^2 \rangle^{(n,m)}$, $\omega^2_\mathrm{nm}$, and $T$, and the unknown effective mass of the mode $\mathrm{m^{(n,m)}_{eff}}$. However,
\new{since $u_{\rm n m}(x,y)=\sin (n\pi x/L) \sin (m\pi y/L)$ is a very good approximation, Eq.~\eqref{realmass} yields}
$\mathrm{m^{(n,m)}_{eff}}=m_T/4$ independent of $(n,m)$, where $m_T$ is the total mass of the membrane. Moreover we expect that for the fundamental mode $\alpha^2_\mathrm{11}(\bar x,\bar y)\simeq 1$ because the measured waist $w = \SI{53.2\pm0.4}{\micro\meter}$ is much smaller than $L= \SI{1}{\milli\meter}$ and the beam is centered very close to the membrane center. As a consequence, we can safely assume $\mathrm{m^{(n,m)}_{eff}} \simeq \mathrm{m^{(1,1)}_{opt}}=\SI{70.0\pm.2}{\nano\gram}$ for the three modes in Eq.~\eqref{alphaesto}, which is also consistent with the value obtained from Eq.~\eqref{realmass} and membrane specifications ($\SI{1x1}{\milli\meter}$ square, $\SI{50}{\nano\meter}$--thick SiN membrane, coated with a $\SI{27}{\nano\meter}$ Nb film with a  $\SI{300}{\micro\meter}$--diameter central circular hole).

\begin{figure}[h]
\centering
\includegraphics[width=0.45\textwidth]{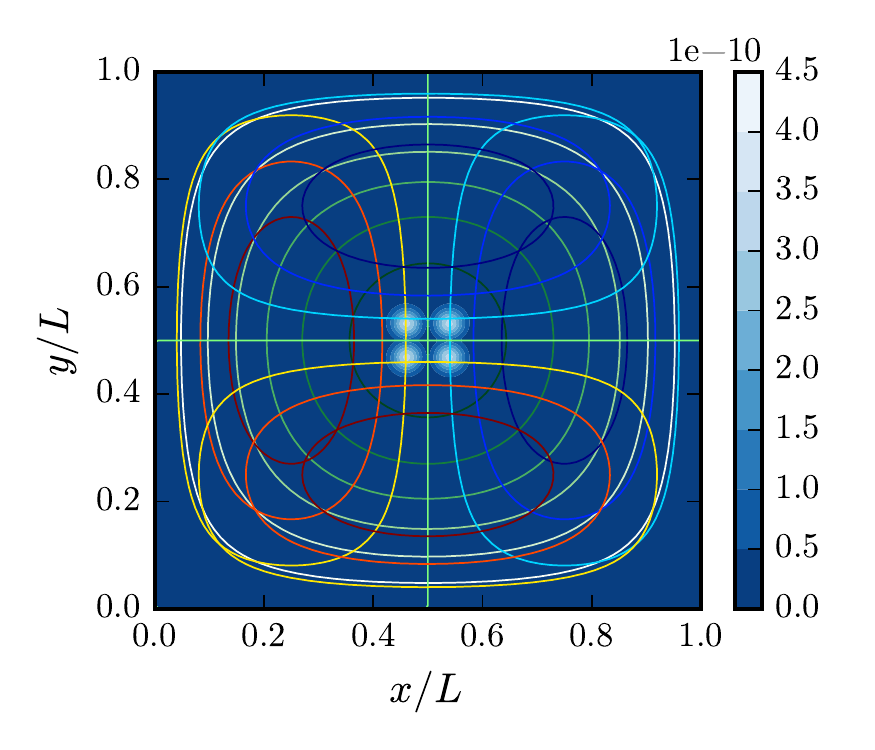}
\caption{Position estimates from the $\chi^2$ minimisation, showing the most likely points.}
\label{fig:lik}
\end{figure}

We then construct the $\chi^2$ quantity
\begin{equation}\label{chisqu}
	\chi^2(x,y) = \sum_{n,m}\big[\alpha_\mathrm{nm}^{\rm{(ex)}}(\bar x,\bar y)  -\alpha_\mathrm{nm}^{\rm{(th)}}(x,y) \big]^2,
\end{equation}
and minimize it over $(x,y)$. The minimizing points $(x_0,y_0)$ are the most likely points, and the corresponding likelihood density function of where the beam is positioned is given by \cite{Nielsen2017}
\begin{equation}
	\mathcal{L}(x,y) = \frac{1}{2\pi\sigma^2}\prod_{n,m}\rm{e}^{-\frac{\big[\alpha_\mathrm{nm}^{\rm{(ex)}}(\bar x,\bar y)  -\alpha_\mathrm{nm}^{\rm{(th)}}(x,y) \big]^2}{2\sigma^2}}
\end{equation}
with $\sigma^2 = \chi^2(x_0,y_0)$, and whose contour plot is shown in Fig.~\ref{fig:lik}. The corresponding best estimation of the transverse overlap for the modes are
\begin{equation}
 	\alpha_\mathrm{11} = \num{0.980}
		\hspace{.3cm}
 	\alpha_\mathrm{12} = \num{0.196}
		\hspace{.3cm}
 	\alpha_\mathrm{21} = \num{0.240}\,,
 \end{equation}
yielding the best estimate for the physical effective mass of the three modes, $\mathrm{m_{eff}^{(1,1)}} \simeq \mathrm{m_{eff}^{(1,2)}} \simeq \mathrm{m_{eff}^{(2,1)}} \simeq \SI{67.3}{\nano\gram}$, within a $3\%$ error, and confirmed by finite element method (FEM) numerical analysis.

\section{The electromechanical couplings}

As shown in Eq.~\eqref{e:G_def} the electromechanical couplings $G_i$ depend upon the explicit expression of the capacitance of the LC circuit and its dependence upon the transverse displacement associated with each vibrational normal mode of the membrane. We can write for the total capacitance $C=C_0+C_{\rm m}(\{x_i\})$, where $C_0$ is the capacitance of the LC circuit (including fixed and tunable capacitors) acting in parallel with the membrane capacitance $C_{\rm m}(\{x_i\})$. We have verified that in our case $C_0 \gg C_{\rm m}(\{x_i\})$, so that from
Eq. (\ref{e:G_def}) one can also write
\begin{equation}\label{eq:coup1}
  G_i \simeq -\frac{V_{\rm dc}}{C_0}\left.\frac{\partial C_{\rm m}(\left\{x_i\right\})}{\partial x_i}\right|_{x_i=\bx_i}.
\end{equation}
Following \cite{Zeuthen2014} and exploiting the geometry of our membrane--electrode arrangement, one can derive a theoretical model of the capacitance $C_{\rm m}(\left\{x_i\right\})$ based on a quasi--electrostatic calculation, which allows to derive both the electromechanical couplings $G_i$ and the frequency shifts of Eq. (\ref{freqshi}), and satisfactorily reproduces the data.

As shown in Fig. 2 of the main text, the membrane capacitor is formed by a four--segment electrode in front of the partially metallized membrane. Since the membrane--electrode separation $h_0$ is significantly smaller than the inter--electrode gaps, we can neglect the direct capacitance between electrode segments; the capacitance is then given by the series of two local contributions, one associated with the positive electrode segments and the membrane in front of it, $C_{+}$, and the second one corresponding to the negative electrode segments, $C_{-}$, i.e.,
\begin{equation}\label{e:C_series_model}
C_{\rm m} = \left[\frac{1}{C_{+}} + \frac{1}{C_{-}}\right]^{-1}.
\end{equation}
For the calculation of $C_{\pm}$ we assume that the curvature of the membrane is sufficiently small so that we can take it to be locally flat. We also neglect edge effects, so that for symmetry, and assuming perfect alignment, we may model the membrane--electrode capacitance locally as that of conducting parallel plates. This local capacitance per area only depends upon the local membrane--electrode separation along the direction normal to the plane defined by the electrodes, and we can write
\begin{equation}\label{localcap}
  C_{\pm}=\int\int d\mathrm{x} d\mathrm{y} \frac{\varepsilon_0\, \xi_{\pm}(x,y)}{h_0+\delta z (x,y)},
\end{equation}
where the integral is taken over the membrane surface, $\xi_{\pm}(x,y)$ is a mask function that equals 1 for points in the membrane plane that are metalized and overlap with the fixed positive or negative electrode, and is zero otherwise, $\delta z (x,y)$ is the membrane displacement field relative to the steady--state configuration, and $\varepsilon_0$ is the vacuum dielectric constant. We can always expand this field in terms of the vibrational eigenmodes $u_{i}(x,y)$ introduced in Eq. (\ref{realmass})
\begin{equation}\label{e:drum_mode_expansion}
\delta z(x,y) = \sum_{i} \beta_{i} u_{i}(x,y),
\end{equation}
where being the eigenmodes $u_{i}(x,y)$ dimensionless, the coefficients $\beta_i$ are canonical drum mode position coordinates. With this notation, the derivatives appearing in the expression for the couplings of Eq. (\ref{eq:coup1}), become
\begin{equation}\label{eq:coup2}
  \left.\frac{\partial C_{\rm m}(\left\{x_i\right\})}{\partial x_i}\right|_{x_i=\bx_i} \rightarrow  \left.\frac{\partial C_{\rm m}}{\partial \beta_i}\right|_{\rm eq},
\end{equation}
where ``eq'' means that the derivative should be evaluated at the static displacement equilibrium configuration of the membrane, $ \delta z (x,y)=0$.
We have explicitly
\begin{equation}\label{chainrule}
  \left.\frac{\partial C_{\rm m}}{\partial \beta_i}\right|_{\rm eq}=\left.\frac{\frac{1}{C_+^2}\left(\frac{\partial C_{+}}{\partial \beta_i}\right) +\frac{1}{C_-^2}\left(\frac{\partial C_{-}}{\partial \beta_i}\right)}{\left(\frac{1}{C_+}+\frac{1}{C_-}\right)^2}\right|_{\rm eq},
\end{equation}
so that, using Eqs. (\ref{localcap})--(\ref{e:drum_mode_expansion}), and inserting the results into Eq. (\ref{eq:coup1}), one finally gets
\begin{equation}\label{e:G_beta_result}
G_i = \frac{V_{\rm DC}\,\epsilon_0}{C_0 h_{0}^2}  A^{\rm eff}_i ,
\end{equation}
where we have defined the effective mode area
\begin{equation}\label{area0}
A^{\rm eff}_i=\left[ \frac{\frac{O^{(1)}_{+,i}}{[O^{(0)}_{+}]^2} + \frac{O^{(1)}_{-,i}}{[O^{(0)}_{-}]^2}}{\left(\frac{1}{O^{(0)}_{+}} + \frac{1}{O^{(0)}_{-}} \right)^2}\right],
\end{equation}
in terms of the quantities
\begin{equation}\label{area1}
O_{\pm,i}^{(j)} \equiv \int\int d\mathrm{x} d\mathrm{y} \xi_{\pm}(x,y)[u_i(x,y)]^j \,\,\,j=0,1.
\end{equation}
The explicit values of the two electromechanical couplings $G_1$ and $G_2$ associated with the two mechanical modes used for our transducer can be obtained from the knowledge of $C_0$, $V_{\rm DC}$, the distance $h_0$ and the various integrals $O_{\pm,i}^{(j)}$. We have evaluated the latter integrals numerically from the calibrated image of the electrode and from the properly normalized finite element numerical solution of the two vibrational eigenmodes, while $C_0$, and $V_{\rm DC}$ are easily measured. The membrane--electrode equilibrium distance $h_0$ instead has been evaluated from the measurement of the mechanical frequency shift of the fundamental vibrational mode.

\subsection{Derivation of the membrane--electrode distance}
\begin{figure}[h!]
\centering
\includegraphics[width=0.4\textwidth]{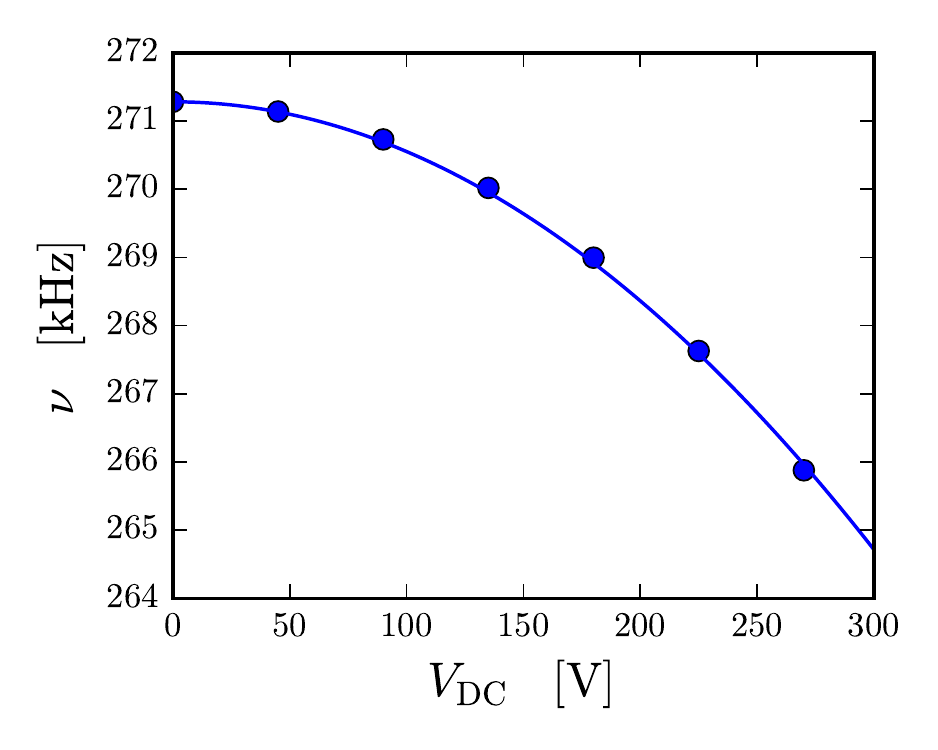}
\caption{Mechanical resonance frequency shift of the fundamental mode as a function of the applied dc voltage $V_{\rm DC}$.}
\label{fig:d0}
\end{figure}

Eq. (\ref{freqshi}) shows that each mechanical mode is shifted quadratically as a function of the applied DC voltage. A measurement of this quadratic phase shift provides a quite accurate indirect method for the determination of the distance $h_0$ between the metalized membrane and the electrode. In our case we have measured the frequency shift of the fundamental mode $(1,1)$ (see Fig. \ref{fig:d0}). Denoting with $i=0$ the fundamental mode (1,1), recalling that $C=C_0+C_{\rm m}(\{x_i\})$ with $C_0 \gg C_{\rm m}(\{x_i\})$ so that $\bar{q} \simeq C_0 V_{\rm DC}$, and using Eq. (\ref{eq:coup2}) and that $\omega_0 = 2\pi \nu_0$, one can rewrite Eq. (\ref{freqshi}) as
\begin{equation}\label{eq:shi2}
  \Delta \nu_0 =-\frac{V_{\rm DC}^2}{16 \pi^2 \mathrm{m_{eff}}\nu_0}
  	\left.\left(\frac{\partial}{\partial \beta_0}\frac{\partial C_{\rm m}}{\partial \beta_0}\right)\right|_{\rm eq},
\end{equation}
where Eq. (\ref{chainrule}) has to be used for the evaluation of $\partial C_{\rm m}/\partial \beta_0$. It is possible to verify that
\begin{equation}\label{approxshi}
  \left.\left(\frac{\partial}{\partial \beta_0}\frac{\partial C_{\rm m}}{\partial \beta_0}\right)\right|_{\rm eq}
  	\simeq \frac{2\varepsilon_0 A^{\rm eff}_0}{h_0^3},
\end{equation}
where $A^{\rm eff}_0$ is the effective area for the fundamental mode, and one can write
\begin{equation}
     \nu(V_{\rm DC}) = \nu_0\Big(1- \frac{\varepsilon_0 A^{\rm eff}_0}{8\pi^2\mathrm{m_{eff}}\nu_0^2 h_0^3}V_{\rm DC}^2\Big)\,.
\end{equation}
On the other hand we can fit the experimental data of Fig. \ref{fig:d0} with
\begin{equation}
     \nu(V_{\rm DC}) = \nu_0\Big(1- \frac{\Lambda}{8\pi^2\mathrm{m_{eff}}\nu_0^2}V_{\rm dc}^2\Big)\,;
\end{equation}
where $\Lambda$ is a fitting parameter. Using the best--fit values derived above, $\nu_0 = \SI{2.71269}{\kilo\hertz}$ and $\mathrm{m_{eff}} = \SI{67.3}{\nano\gram}$, the best--fit value of the parameter $\Lambda=\SI{105.2 \pm 0.9}{\micro\farad\per\square\meter}$, and using $A^{\rm eff}_0 = \SI{0.3546}{\square\milli\meter}$, and $\epsilon_0 = \SI{8.854e-12}{\second^4\per\meter^3\per\kilogram}$, the distance between membrane and electrode is evaluated to be
\begin{equation}\label{eqho}
 	h_0 =\left(\frac{\varepsilon_0 A^{\rm eff}_0}{\Lambda}\right)^{\frac{1}{3}}\simeq \SI{31.0 \pm 0.1}{\micro\meter}\,.
\end{equation}

\begin{figure}[h!]
\includegraphics[width=0.4\textwidth]{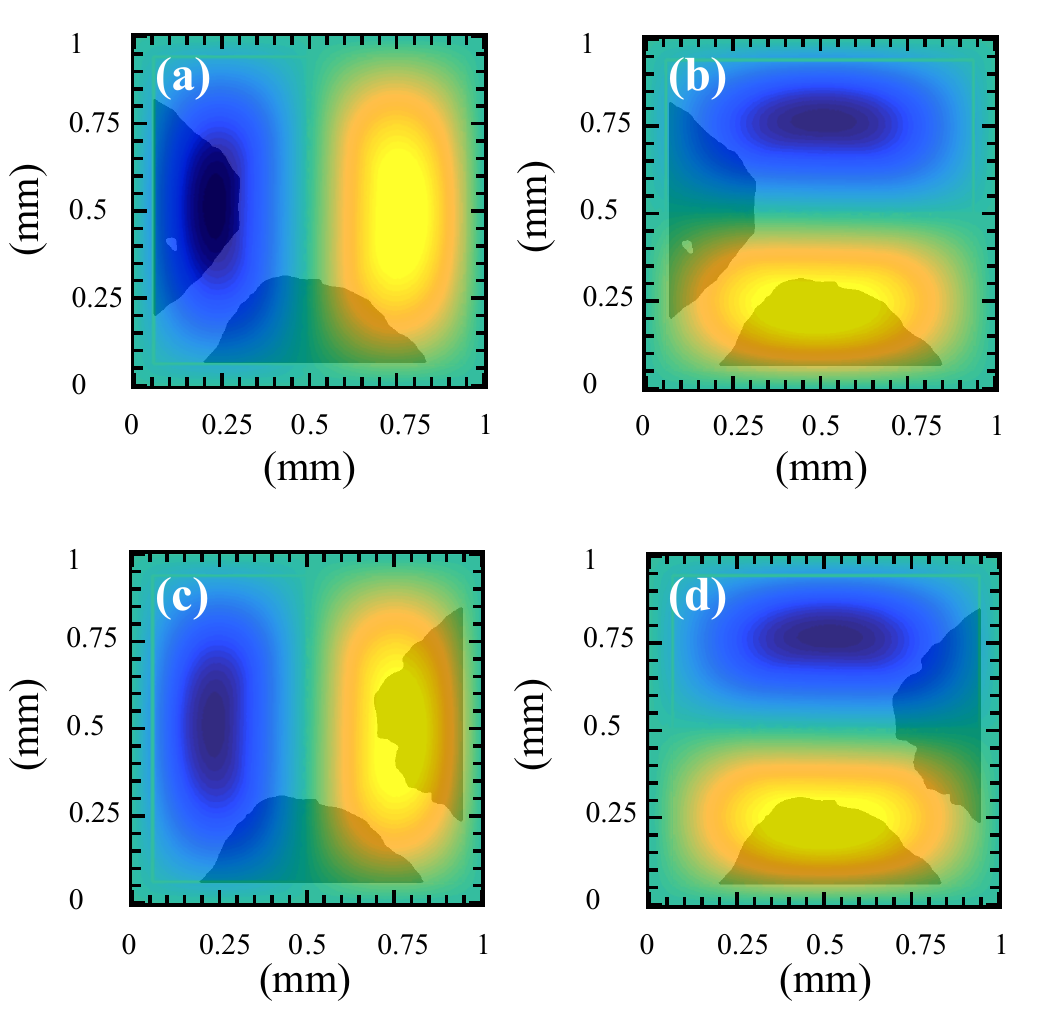}
\caption{
\new{Image (grey shape) of the electrodes used for realizing constructive interference, (a) and (b),  and for destructive interference, (c) and (d), overlapped with the mode shapes of the doublet obtained from finite element simulation. From these images we have calculated the effective areas defined in Eq.~(\protect\ref{area0}) and obtained $G_1=\SI{116.4}{\volt\per\meter}$ and $G_2=\SI{-109.6}{\volt\per\meter}$ for the upper electrode configurations, and $G_1=\SI{117.0}{\volt\per\meter}$ and $G_2=\SI{113.9}{\volt\per\meter}$ for the lower electrode configurations. These values are in very good agreement with the values given in the main text and obtained as best-fit parameters of the measured output spectra}.
} \label{modeelectrod}
\end{figure}

With this derivation of the membrane--electrode distance $h_0$, we can finally estimate the electromechanical couplings $G_1$ and $G_2$ using Eq. (\ref{e:G_beta_result}) once that the effective areas $A^{\rm eff}_i$ have been estimated using Eqs.~(\ref{area0})--(\ref{area1}).
For the mode--electrode configuration of Fig.~(\ref{modeelectrod}), our numerical estimation gives the values of $-0.0178$ mm$^2$ for Fig.~(\ref{modeelectrod}a), $0.0189 $ mm$^2$ for Fig.~(\ref{modeelectrod}b), $0.0185$ mm$^2$ for Fig.~(\ref{modeelectrod}c), and $0.0190$  mm$^2$ for Fig.~(\ref{modeelectrod}d).
\new{These values of the effective area can be understood from the fact that the blue and yellow lobes denote respectively the negative and positive parts of the vibrational mode function. In each of the four configurations one of the two electrodes has approximately the same overlap with the positive and negative lobes, yielding therefore a negligible contribution to the effective area of Eq. (\ref{area0}). The other electrode yields the main contribution to the effective area which is therefore negative for Fig.~(\ref{modeelectrod}a) and positive for the other three cases, so that the upper configurations corresponds to the constructive interference case and the lower ones to the destructive interference case}.

Using these values for the effective areas, and inserting
$ C_0=404 $ pF, $V_{dc} = 270 $ V and $h_{0}= \SI{31.0}{\micro\meter}$ into Eq. (\ref{e:G_beta_result}), we get $G_1=\SI{116.4}{\volt\per\meter}$ and $G_2=\SI{-109.6}{\volt\per\meter}$ for the upper electrode configurations corresponding to the constructive interference case (see Fig.~\ref{modeelectrod}a and \ref{modeelectrod}b). Instead we get $G_1=\SI{117.0}{\volt\per\meter}$ and $G_2=\SI{113.9}{\volt\per\meter}$ for the lower electrode configurations corresponding to the destructive interference case (see Fig.~\ref{modeelectrod}c and Fig.~\ref{modeelectrod}d).
\new{These values are in very good agreement with the values given in the main text and obtained as best-fit parameters of the measured output spectra}.

\begin{thebibliography}{99}
\bibitem{RMP2014} M. Aspelmeyer, et al., Rev. Mod. Phys. {\bf 86}, 1391 (2014).
\bibitem{OConnell2010} A. D. O’Connell, et al., Nature {\bf 464}, 697 (2010).
\bibitem{Teufel2011a} J. D. Teufel, et al., Nature {\bf 475}, 359 (2011).
\bibitem{Chan2011} J. Chan, et al., Nature {\bf 478}, 89 (2011).
\bibitem{Verhagen2012} E. Verhagen, et al., Nature {\bf 482}, 63 (2012).
\bibitem{Safavi-Naeini2013} A. H. Safavi-Naeini, et al., Nature {\bf 500}, 185 (2013).
\bibitem{Palomaki2013a}  T. A. Palomaki, et al., Science {\bf 342}, 710 (2013).
\bibitem{Peterson2016} R. Peterson, et al., Phys. Rev. Lett. {\bf 116}, 063601
(2016).
\bibitem{Clark2016} J. B. Clark, et al., Nat. Phys. {\bf 12}, 683 (2016).
\bibitem{Riedinger2016} R. Riedinger, et al., Nature {\bf 530}, 313 (2016).
\bibitem{Pikovski2012} I. Pikovski, et al., Nat. Phys. {\bf 8}, 393 (2012).
\bibitem{Bawaj2015} M. Bawaj, et al., Nat. Commun. {\bf 6}, 7503 (2015).
\bibitem{Chaste2012} J. Chaste, et al., Nature Nanotech. {\bf 7}, 301 (2012).
\bibitem{Moller2017} C. B. Møller, et al., Nature {\bf 547}, 191 (2017).
\bibitem{Stannigel2010} K. Stannigel, et al., Phys. Rev. Lett. {\bf 105}, 220501 (2010).
\bibitem{Regal2011} C. A. Regal, et al., J. Phys. Conf. Ser. {\bf 264}, 012025 (2011).
\bibitem{Taylor2011}  J. M. Taylor, et al., Phys. Rev. Lett. {\bf 107}, 273601 (2011).
\bibitem{Wang2012}  Y.-D. Wang, et al., Phys. Rev. Lett. {\bf 108}, 153603 (2012).
\bibitem{Tian2012} L. Tian, Phys. Rev. Lett. {\bf 108}, 153604 (2012).
\bibitem{Barzanjeh2012} S. Barzanjeh, et al., Phys. Rev. Lett. {\bf 109}, 130503 (2012).
\bibitem{Bagci2013} T. Bagci, et al., Nature {\bf 507}, 81 (2013).
\bibitem{Andrews2014} R. W. Andrews, et al., Nat. Phys. {\bf 10}, 321 (2014).
\bibitem{Fink2016} J. M. Fink, et al., Nat. Commun. {\bf 7}, 12396 (2016).
\bibitem{NMR}K. Takeda, et al.,  arXiv:1706.00532 [quant-ph] (2017).
\bibitem{Zeuthen2016} E. Zeuthen, et al., arXiv:1610.01099 [quant-ph] (2016).
\bibitem{Nielsen2017} W. H. P. Nielsen, et al., Proc. Natl. Acad. Sci. (USA) \textbf{144}, 62
(2017).
\bibitem{Massel2012}F. Massel, et al., Nat. Commun. \textbf{3}, 987 (2012).
\bibitem{Noguchi2016}A. Noguchi, et al., New J. Phys. \textbf{18}, 103036 (2016)
\bibitem{Han2016}X. Han, et al., Phys. Rev. Lett. {\bf 117}, 123603 (2016).
\bibitem{Caniard2007} T. Caniard, et al., Phys. Rev. Lett. {\bf 99}, 110801 (2007).
\bibitem{Cerdonio2001} M. Cerdonio, et al., Phys. Rev. Lett. {\bf 87}, 031101 (2001).
\bibitem{Xu2016} X. W. Xu, et al., Phys. Rev. A {\bf 93}, 023827 (2016).
\bibitem{Bernier2016} N. R. Bernier, et al., Nat. Commun. \textbf{8}, 604 (2017)
\bibitem{Peano2015} V. Peano, et al., Phys. Rev. X {\bf 5}, 031011 (2015).
\bibitem{Peterson2017} G. A. Peterson, et al., Phys. Rev. X \textbf{7}, 031001 (2017)
\bibitem{OurNote} Control of interference phenomena in multimode optomechanical systems for the realisation of nonreciprocal devices through the control of the driving phases has been
exploited also in Refs. \cite{Fang2017,Li2017,Kuzyk2017}, which are however based on configurations different from the one demonstrated here.
\bibitem{Fang2017} K. Fang, et al., Nat. Phys. {\bf 13}, 465 (2017).
\bibitem{Li2017} Y. Li, et al., arXiv:1705.08635  [quant-ph] (2017).
\bibitem{Kuzyk2017}M. C. Kuzyk and H. Wang, arXiv1705.04722 (2017).
\bibitem{Biancofiore2011}C. Biancofiore, et al., Phys. Rev. A
\textbf{84}, 033814 (2011).
\bibitem{Zeuthen2014}E. Zeuthen, Ph.D. thesis, University of Copenhagen (2014).
\end{thebibliography}
\end{document}